\documentclass[fleqn,usenatbib]{mnras}

\usepackage{newtxtext,newtxmath}

\usepackage[T1]{fontenc}
\usepackage{orcidlink}

\DeclareRobustCommand{\VAN}[3]{#2}
\let\VANthebibliography\thebibliography
\def\thebibliography{\DeclareRobustCommand{\VAN}[3]{##3}\VANthebibliography}

\usepackage{graphicx}	
\usepackage{amsmath}	
\usepackage{hyperref}

\newcommand{\oat}{\ensuremath{o_{\rm ATLAS}}}
\newcommand{\cat}{\ensuremath{c_{\rm ATLAS}}}
\newcommand{\gps}{\ensuremath{g_{\rm PS}}}
\newcommand{\rps}{\ensuremath{r_{\rm PS}}}
\newcommand{\ips}{\ensuremath{i_{\rm PS}}}
\newcommand{\zps}{\ensuremath{z_{\rm PS}}}
\newcommand{\yps}{\ensuremath{y_{\rm PS}}}
\newcommand{\wps}{\ensuremath{w_{\rm PS}}}
\newcommand{\grips}{\ensuremath{gri_{\rm PS}}}
\newcommand{\rizps}{\ensuremath{riz_{\rm PS}}}
\newcommand{\grizyps}{\ensuremath{grizy_{\rm PS}}}

\newcommand{\degree}{\mbox{$^\circ$}}

\newcommand{\kms}{\mbox{$\rm{km}\,s^{-1}$}}

\newcommand{\HO}{70} 
\newcommand{\zlim}{0.045} 
\newcommand{\dlim}{200} 
\newcommand{\Rlim}{50} 

\title[Contamination by Massive Stellar Outbursts]{Results from the Pan-STARRS Search for Kilonovae: Contamination by Massive Stellar Outbursts}

\author[M. D. Fulton et al.]{
M. D. Fulton,$^{1}$\thanks{E-mail: mfulton07@qub.ac.uk}\orcidlink{0000-0003-1916-0664},
S. J. Smartt,$^{2,1}$\orcidlink{0000-0002-8229-1731},
M. E. Huber$^{3}$\orcidlink{0000-0003-1059-9603},
K. W. Smith$^{2,1}$\orcidlink{0000-0001-9535-3199},
K. C. Chambers$^{3}$\orcidlink{0000-0001-6965-7789},
M. Nicholl$^{1}$\orcidlink{0000-0002-2555-3192},
\newauthor{
S. Srivastav$^{2}$\orcidlink{0000-0003-4524-6883},
D. R. Young$^{1}$\orcidlink{0000-0002-1229-2499},
E. A. Magnier$^{3}$\orcidlink{0000-0002-7965-2815},
C.-C. Lin$^{3}$\orcidlink{0000-0002-7272-5129},
P. Minguez$^{3}$,
T. de Boer$^{3}$,
T. Lowe$^{3}$,
}\newauthor{
R. Wainscoat$^{3}$\orcidlink{0000-0002-1341-0952}
}
\\\\
$^{1}$Astrophysics Research Centre, School of Mathematics and Physics, Queen's University Belfast, BT7 1NN, UK\\
$^{2}$Department of Physics, University of Oxford, Keble Road, Oxford, OX1 3RH, UK\\
$^{3}$Institute for Astronomy, University of Hawai'i, 2680 Woodlawn Drive, Honolulu, HI 96822, USA\\
}

\date{Submitted to MNRAS May-2025.}

\pubyear{2025}

\begin{document}
\label{firstpage}
\pagerange{\pageref{firstpage}--\pageref{lastpage}}
\maketitle

\begin{abstract}
We present results from the Pan-STARRS optical search for kilonovae without the aid of gravitational wave and gamma-ray burst triggers. 
The search was conducted from 26 October 2019 to 15 December 2022. 
During this time, we reported 29,740 transients observed by Pan-STARRS to the IAU Transient Name Server. 
Of these, 175 were Pan-STARRS credited discoveries that had a host galaxy within 200\,Mpc and had discovery absolute magnitudes $M>-16.5$. 
A subset of 11 transients was plausibly identified as kilonova candidates by our kilonova prediction algorithm. 
Through a combination of historical forced photometry, extensive follow-up, and aggregating observations from multiple sky surveys, we eliminated all as kilonova candidates. 
Rapidly evolving outbursts from massive stars (likely to be Luminous Blue Variable eruptions) accounted for 55\% of the subset's contaminating sources. 
We estimate the rate of such eruptions using the ATLAS $100$\,Mpc volume-limited survey data. 
As these outbursts appear to be significant contaminants in kilonova searches, we estimate contaminating numbers when searching gravitational wave skymaps produced by the LIGO-Virgo-Kagra science collaboration during the Rubin era. 
The Legacy Survey of Space and time, reaching limiting magnitudes of $m\approx25$, could detect 2-6 massive stellar outbursts per $500$\,deg$^{2}$ within a 4-day observing window, within the skymaps and volumes typical for binary neutron star mergers projected for Ligo-Virgo-Kagra Observing run 5. 
We conclude that while they may be a contaminant, they can be photometrically identified. 

\end{abstract}

\begin{keywords}
editorials, notices ---
surveys ---
techniques: photometric ---
stars: variables: S Doradus ---
(transients:) neutron star mergers ---
transients: supernovae
\end{keywords}



\section{Introduction}
\label{sec:intro}
The historic gravitational wave (GW) event GW170817 resulting from a binary neutron star (BNS) merger \citep{2017PhRvL.119p1101A} produced a short gamma-ray burst (sGRB) GRB170817A \citep{2017ApJ...848L..13A} and a rapidly evolving optical and infrared transient AT~2017gfo \citep{MMApaper2017}. 
A distance estimate from LIGO-Virgo of 40$^{+8}_{-14}$\,Mpc provided a 3-dimensional volume small enough to allow the technique of targeting specific, catalogued galaxies with redshifts matching the distance estimate to locate either an X-ray, ultraviolet (UV), optical or near-infrared (NIR) transient to arcsecond precision. 
This was achieved just 11\,hours later, during the first night in Chile, when the object was discovered in optical and NIR images \citep{Arcavi2017, Coulter2017, Lipunov17, SoaresSantos2017, Tanvir2017, Valenti2017}. 
Global monitoring followed, with the spectra from Chile 24 hours later showing an unprecedented evolution and confirming that this was the spectral signature of a unique transient with no known counterpart \citep{Chornock2017, McCully2017, Nicholl2017, 2017Natur.551...67P, Shappee2017, Smartt2017}. 

Extensive data collected showed the transient to fade rapidly while becoming significantly redder, such that most of the flux after one week was emitted in the NIR, and possibly even beyond 2.5\,$\mu$m after 8$-$10 days \citep{Andreoni2017, Cowperthwaite2017, Drout2017, Kasliwal2017, Evans2017, Smartt2017, Tanvir2017, Troja2017, Utsumi2017}. 
Light curve and spectral modelling showed that AT~2017gfo was the first confirmed candidate for a kilonova, powered by radioactive decay of neutron-rich material synthesised in the merger \citep{1998ApJ...507L..59L, 2010MNRAS.406.2650M, 2013ApJ...774...25K, 2013ApJ...775..113T, 2017CQGra..34j4001R, Kasen2017}. 
With this extraordinary transient undoubtedly associated with the gravitational wave event, a new era of multi-messenger astronomy was born. 
The third LIGO-Virgo science run (O3) produced detections of a second high-significance BNS merger \citep[GW190425]{GW190425}, an event with a secondary component mass at the limit of stable neutron stars \citep[GW190814]{GW190814} and two black hole - neutron star mergers \citep[GW200105 \& GW200115]{LVC_O3_NSBHs}. 
Despite efforts across many wavelengths, no plausible electromagnetic counterparts were identified for any of these three events \citep{2020ApJ...895...96V,2021A&A...650A.131B,2021ApJ...923...66A,2021ApJ...923..258K,2021ApJ...923L..32D,190425_panstarrs}. 

Searches for kilonovae without a GW or GRB trigger have been ongoing for some years, and the observed light curve of AT~2017gfo has instilled hope that such events would be detectable within 200\,Mpc. 
Wide-field survey strategies adopted by the likes of the All-Sky Automated Survey for Supernovae \citep[ASAS-SN;][]{2014ApJ...788...48S}, the Asteroid Terrestrial-impact Last Alert System \citep[ATLAS;][]{2018PASP..130f4505T,2020PASP..132h5002S}, the Pan-STARRS Survey for Transients \citep[PSST;][]{2015IAUGA..2258303H} and the Zwicky Transient Facility \citep[ZTF;][]{2019PASP..131a8002B} have resulted in the discovery of various, fast-evolving transients, defined roughly as displaying thermal emission that evolves on timescales of days and significantly faster than the principal Type Ia, Type Ibc and Type II supernova populations. 

Fast blue optical transients (FBOTs) such as AT~2018cow \citep{2018ATel11727....1S, 2018ApJ...865L...3P, 2019ApJ...872...18M,2019MNRAS.484.1031P}, ZTF18abvkwla \citep{2020ApJ...895...49H}, CSS161010 \citep{2020ApJ...895L..23C} and AT~2020xnd \citep{2021MNRAS.508.5138P,2022ApJ...932..116H} have been discovered by optical surveys and found to emit from hard X-rays to radio. 
AT~2018kzr \citep{2019ApJ...885L..23M} is the fastest optical transient (apart from relativistic, jet-driven transients) discovered since AT~2017gfo and may also be a compact binary merger \citep{2020MNRAS.497..246G}. 
The faint Ca-rich SN~2019bkc and ultra-stripped SN~2019wxt are also remarkably rapid in their evolution \citep{chen2020most,prentice2020rise,2023A&A...675A.201A}. 
The recent discovery of SN~2022aedm by ATLAS has proposed a new class of luminous, fast-cooling transients (LFCs) whose luminosity and rapid evolution rival that of FBOTs \citep{Nicholl_2023aedm}.

However, even with the ASAS-SN, ATLAS and ZTF surveys now working at cadences of 1$-$2 days across their respective visible skies and the ability to combine the data from all, the detection rates of these fast-evolving transients remain relatively low, indicating low volumetric rates. 
An archival search of the Pan-STARRS survey data identified two fast-fading transients, PS15cey \& PS17cke, whose origins are uncertain but are possible kilonova candidates \citep{2021MNRAS.500.4213M}. 
The ZTF Realtime Search and Triggering (ZTF-ReST) project \citep{Andreoni2021fast} designed to identify kilonova candidates in near real-time has been successful in picking out numerous GRB afterglows and fast-fading transients within the ZTF data-stream: AT~2020sev \citep{Andreoni2020zwicky}, AT~2020yxz \citep{coughlin2020ztf20acozryr}, AT~2021buv \citep{kool2021ztf21aagwbjr}, and AT~2021clk \citep{andreoni2021ztf21aahifke}. 
However, no kilonovae were found and a rate limit of $R < 900$ Gpc$^{-3}$yr$^{-1}$ was derived \citep{Andreoni2020zwicky} for AT~2017gfo-like transients. 
We currently run a real-time filter on Lasair broker's processing of the ZTF public stream \citep{2024RASTI...3..362W} that flags fast-evolving candidates \citep[FastFinder;][]{2023TNSAN.339....1F} and have found no plausible kilonovae. 
Searches with ATLAS \citep{Srivastav2022} and DECam \citep{2025MNRAS.537.3332V} have also found no plausible kilonova candidates amongst the intrinsically faint transient population they probed. 

A focused search of the long-duration GRB~211211A \citep{mangan2021grb} revealed an optical afterglow candidate \citep{zheng2021grb} whose luminosity, duration and colour are comparable with that observed in AT~2017gfo, confirming the detection of a kilonova as shown by \cite{2022Natur.612..223R}.
The second brightest GRB ever detected \citep[GRB~230307A;][]{2023GCN.33411....1D,2023ApJ...954L..29D} also produced an optical afterglow that was followed by a thermal emission component consistent with radioactively powered kilonova emission \citep{2024Natur.626..737L,2024Natur.626..742Y}.
James Webb Space Telescope spectra of this afterglow have revealed the presence of tellurium in the spectra \citep{2024Natur.626..737L,2025MNRAS.538.1663G}, confirming the existence of another kilonova connected to a long GRB. 
These recent discoveries have contradicted the widely accepted theory that long GRBs are only caused by supernovae. 
Nevertheless, no convincing kilonova transient has been found in any sky survey search without the aid of either a GW or GRB trigger.

The paper outline is as follows: Section\,\ref{sec:search} reviews the Pan-STARRS telescope system, data reduction and the strategies adopted in the search for kilonovae project. 
Section\,\ref{sec:results} illustrates the search results, the deliberation provided for all kilonova candidates identified and their subsequent elimination as candidates. 
Section\,\ref{sec:disc} describes the primary contaminants predicted for present and future kilonova searches and analyses of the LBV contamination rate. 
Finally, Section\,\ref{sec:conclusions} summarises this paper. 
Throughout the paper, we adopt a Hubble constant of $H_{0}=$\HO$\pm5$\kms, which brackets the current range of values from both local Cepheid and type Ia supernova measurements and the CMB \citep{Riess2019, Planck2016Cosmo}. 
A flat cosmology is assumed ($\Lambda$CDM cosmology with $\Omega_{\rm M} = $0.3), but at low redshifts, the values of $\Omega_{\rm M}$ and $\Omega_{\rm \Lambda}$ are not relevant to distance calculations at our required precision.

\section{The Pan-STARRS search for kilonovae}
\label{sec:search}

\subsection{The Pan-STARRS Telescope System} 
\label{subsec:tele-system}
The Pan-STARRS system comprises two 1.8\,m telescope units located at the summit of Haleakala on the Hawaiian island of Maui \citep{2016arXiv161205560C}. 
The first telescope, Pan-STARRS1 (PS1), commenced operations in 2010 and is fitted with a 1.4 Gigapixel camera (GPC1) with 0.26\arcsec pixels providing a focal plane of 3.0-degree diameter, which corresponds to a field of view area of 7.06 square degrees. 
The second telescope, Pan-STARRS2 (PS2), fully commissioned in 2019, is fitted with a similar but larger 1.5 Gigapixel camera (GPC2), resulting in a slightly wider field of view. 
Both telescopes are equipped with an SDSS-like filter system, denoted as \grizyps and a broad \wps\,passband that is a composite of \gps+\rps+\ips\,\citep{2012ApJ...745...42T}.

The Pan-STARRS1 Science Consortium (PS1SC) 3$\pi$ Survey produced \grizyps\,images of the whole sky north of $\delta = $-30$^{\circ}$ \citep{2016arXiv161205560C}. 
Multi-epoch observations spanning 2009$-$2014 have been stacked, and a public data release provides access to the images and catalogues \citep{flewelling2020}. 
We also have proprietary \ips\ coverage between -40$^{\circ} < \delta < $-30$^{\circ}$, and coverage of most of the northern sky in \wps. 
The \wps-stack has been built up through the Near-Earth Object (NEO) surveys and can reach depths of \wps$\simeq$24. 
These data provide reference images for immediate template subtraction of new images, which allows the discovery of transients and accurate photometry with host galaxy light removed.

Since early 2014, Pan-STARRS has primarily been funded by the Near Earth Observation program of the National Aeronautics and Space Administration to search for NEOs with additional contributions from collaborative transient programs. 
Pan-STARRS, now operated by the Pan-STARRS New Science Consortium, is the leading NEO discovery survey and discovers more than half of the larger (>140-metre diameter) NEOs \citep{wainscoat2019science}. 

\subsection{Telescope scheduling and cadence} 
\label{subsec:observe-strat}

The fields visited and the overall cadence of the Pan-STARRS observations on a particular region of the sky are optimised for discovering NEOs, not routine observing of stationary transients.
In normal survey mode, Pan-STARRS takes 4$\times$45s exposures in \wps\,during dark time, or a combination of \ips,\, \zps\,and \yps\, during bright time, reaching typical 5$\sigma$ magnitudes of \wps<22, \ips<21.5, \zps<20.9, and \yps<19.7 per individual exposure
\footnote{The silver coating on the secondary mirror of PS2 started degrading in mid-2019 (believed due to volcanic gases from the Kilauea eruption prior) and before re-coating in mid-2021 the sensitivity had diminished by $\sim$0.9 magnitudes.} 
\citep{2015IAUGA..2258303H, 2016arXiv161205560C}. 
For every quad exposure set, the return visits are co-aligned and spaced 10$-$20 minutes apart to be able to link and identify moving objects \citep{2015IAUGA..2251124W,wainscoat2019science,2022DPS....5450401W}. 

A collection of quads make up a chunk that generally spans approximately one hour of Right Ascension within a 10-degree Declination band.
Sets of chunks on the sky are scheduled nightly to cover areas that have a greater likelihood for NEO discovery or have not been observed during the current lunation while also avoiding sky limitations like the Moon (30-degree avoidance angle), the Galactic plane and bulge (10 and 15 degrees respectively), the keyhole and the DKIST facility on the southern horizon along with alternate chunks to accommodate poorer seeing in particular areas of the sky from prevailing winds. 
NEOs discovered during any night will also have additional follow-up observations later that same night or the following night to improve the orbital solution. 
Depending on the speed of the NEO and the delay in the return visit, these observations may have some fraction of overlap with the original quad field.
Generally, the goal is to cover the entire sky visible by Pan-STARRS at least once per lunation, and the closer the field is to the ecliptic, the higher the probability of there being more than one night of observation on that field per lunation.
Fields having multiple nights of observations within a lunation, and in other lunations, are all done on a different sky tessellation (orientation on the sky) to fill in the cell and chip gaps of the detector for deeper static sky image products, and so may not always have pixel coverage on a specific overlapping region in all return visits. 
The fill factor of the GPC1 and GPC2 cameras is of order 60-76\% \citep{2016arXiv161205560C}. 
During bright time, the \ips\ (Moon illumination greater than 60\%) and \zps\ (one day around full Moon) chunks are more focused in the northern celestial hemisphere, farther from the ecliptic, to build up inputs for deep static sky stacks.
Twilight for \zps, \yps\ filters at the start and end of the night are too short for quad sets for NEO discovery and are used instead to also contribute to deeper static sky stacks and can have upwards of 10 repeat nightly observations on a particular field in a lunation (albeit also on different tessellations).
Effectively, the current NEO search survey, with more of a focus on the \wps, \ips\ filters and irregular cadence, is treated similarly to the original PS1SC 3$\pi$ survey with the data similarly processed.

\subsection{Data Processing and search for transients and kilonovae} 
\label{subsec:dataproc}

We have been carrying out the Pan-STARRS Search for Transients \citep[PSST;][]{huber2017} since 2017 in the ongoing data from the Pan-STARRS NEO survey. The basis of this programme is the discovery of extragalactic transients combined with sampling of their lightcurves by the NEO survey data and dedicated follow-up by the two Pan-STARRS telescopes. 
The "Pan-STARRS Search for Kilonovae" programme \citep{smartt2019pssk}, is an offshoot of the PSST and is a focused search for intrinsically faint and rapidly evolving transients coincident with galaxies that are within 200\,Mpc, according to either spectroscopic redshifts or direct distance estimates.

Each image the telescopes take is processed on the summit and written to disk on a collection of computers in the Pan-STARRS facility called “pixel servers". 
Shortly after the images are written, a summary listing of the image and exposure information is added to the summit datastore. 
During the night, the Pan-STARRS Image Processing Pipeline (IPP) regularly monitors the datastore for new images that are ready for processing. 
The IPP convolves and subtracts archive reference images of the same sky area from the newly acquired data. 
Difference images and catalogues (FITS tables) of difference image sources are produced and published to the Manoa datastore. 
For more details on the pipeline processes, see \cite{Magnier2020IPP}. 
This datastore is regularly monitored by software at QUB, which downloads the catalogues (via HTTP) ready to be ingested into an extensive MySQL database. 
Difference detections in the FITS tables are annotated with bitmasks that can identify certain image subtraction artefacts. 
Detections that fail this flag test and are within $\pm$5 degrees of galactic latitude are immediately rejected. 
The rest are ingested, and detections co-located in the sky (within 1\arcsec) are aggregated into difference objects in the database.

Once the data are inserted, post-ingest filters are applied to the newly created (or appended) objects. 
The initial post-ingest filter requires at least two `quality' detections for each aggregated object. 
These detections must not be flagged in the additional crosstalk bitmasks, and must be at least 5$\sigma$ significance. 
After this, the catalogued measurements of each detection in the database (mostly relating to PSF fitting) associated with this subset of objects is passed through a random forest machine learning algorithm and given a real-bogus score in catalogue space. 
Those with a median value below a fixed threshold are rejected. 

We then cross-match the potential, astrophysical, real transients against a large-scale, custom-built master catalogue of galaxies using the contextual transient classifier "Sherlock" as described in \cite{2020PASP..132h5002S}. 
Sherlock mines a custom-built library of historical and ongoing astronomical survey data and attempts to predict the nature of a transient based on the resulting crossmatched associations found. 
This serves to identify bright and variable stars, which we archive within our local Pan-STARRS transient server. 
For the surviving objects, the most recent (up to) six difference image stamps are requested from the postage stamp server in Hawaii. 
After the stamps are downloaded, a convolutional neural network, trained to differentiate PSF-like sources (e.g. previously human-scanned SNe and known slow-moving objects) from non-PSF-like sources, is run on the $20\times20$ pixel core of the stamps, which are centred on the transient \citep[similar to that described in][]{Wright15,2015MNRAS.449..451W,2024RASTI...3..385W}.
The resulting median real-bogus score for each object is generated, and those below a set threshold are rejected.
Sherlock's library of galaxy catalogues contains photometric and spectroscopic redshifts, and Sherlock can remotely query the \citet{ned1} to acquire galaxy luminosity distance information and also determine projected galactocentric distances, which are used in the selection of candidates for the kilonovae search. 
All remaining objects are then locally crossmatched with a database of solar system objects to further locate contaminants (e.g. slow-moving asteroids), which are moved aside, leaving candidate extragalactic transients associated with potential host galaxies or `orphans' to be scanned and vetted manually.
Nuclear transients and known AGN are also flagged if the source in the difference image is within 1.5\arcsec of the centre of a galaxy. 

The Pan-STARRS Search for Kilonovae builds on this search for transients by focusing on the local Universe - galaxies with known redshifts and intrinsically faint transients associated with these. To summarise, potential transients of interest to the kilonovae search are identified and flagged if they pass the following criteria 

\begin{itemize}
    \item $N\ge2$ number of detections temporally separated by at least 10 minutes (each with minimum 5$\sigma$ significance) in any one night and within 1\arcsec of each other. 
    \item All detections must successfully pass through the IPP. 
    \item A CNN machine learning algorithm, similar to that described in \cite{2024RASTI...3..385W}, provides a real-bogus score based on the pixel quality for all detections on the first epoch. The average of these must be above 0.67. 
    \item The transient must be contextually classified as extragalactic and associated with a potential host galaxy by the Sherlock algorithm.
    \item The transient must fall within a projected galactocentric distance of $R_{\rm g}\le$\Rlim\,Kpc from the associated host.
    \item The associated host must have a luminosity distance $D_{\rm L}\leq\dlim$\,Mpc. This can come from either a redshift-independent direct distance estimate or a redshift-derived distance measurement.
\end{itemize}

These last two cuts arise from the likely nature of the optical counterparts of BNS systems and the practical limitations of our follow-up response.
\cite{2021ApJ...909..207R} and \cite{10.1093/mnras/stac1473} predict that for low redshifts ($z\leq$0.05), the majority of BNS mergers will be found in star-forming hosts, and the lifetimes of such progenitors systems are short ($<<$1\,Gyr; \citealt{Beniamini_2016,10.1093/mnras/stac1473}). 
Such systems prefer relatively low natal-kick velocities ($v_{k}$<100\kms; \citealt{10.1093/mnras/sty1999,10.1093/mnras/stac1473}), and most can be found within 10\,Kpc of their host. 
Longer-lived systems ($\geq$1\,Gyr) are also possible \citep{10.1093/mnras/stac1473} in low redshift galaxies and have the potential to travel beyond 100\,Kpc, as long as the natal kick velocities are sufficient to escape their host galaxy. 
Consequently, transients discovered at such extreme apparent galactocentric distances will often be unrelated sources, appearing close to these galaxies in projection because of chance coincidence. 
Such transients are more likely associated with foreground stellar eruptions, such as Cataclysmic Variables (CVs), or distant supernovae with unresolved hosts. 
Therefore, we adopted a practical limit of $R_g\leq$\Rlim\,Kpc to maximise the recovery of kilonovae while reducing the contamination of such chance coincidences.

To measure a genuine change in brightness and thus investigate a transient for kilonova-like characteristics, we require all follow-up observations to have a sufficient signal-to-noise ratio. 
The sensitivity of the Pan-STARRS system, the speed at which follow-up can occur after discovery, and the rate at which kilonovae are expected to fade dictate the limits of our search. 
The time between the first detections of a transient in Pan-STARRS, data processing, a human recognising the nature of the transient on the Pan-STARRS science server, scheduling follow-up, and finally obtaining the additional data typically exceeds 36 hours. 
Within a 36$-$48 hour period, we expect a kilonova candidate to fade by 0.5$-$1.0 magnitudes, depending on the filter. We estimated this from a compiled dataset of the original photometry of AT~2017gfo \citep{Arcavi2017,Smartt2017,Tanvir2017,Andreoni2017,Cowperthwaite2017,Drout2017,Kasliwal2017,Evans2017,Troja2017,Utsumi2017} and we constructed a synthetic \wps\,filter light curve. 
The measured mean decline rates are 0.52\,mag\,d$^{-1}$, 0.49\,mag\,d$^{-1}$ and 0.37\,mag\,d$^{-1}$ for the \wps,\,\ips\,and \zps\,filters respectively over the 14-day lifetime of the event. 
AT~2017gfo reached a peak absolute magnitude of $M\simeq$-15.8 (AB mag) in the $griz$ optical filters. 
Various attempts to synthesise theoretical light curves of BNS mergers \citep{2015MNRAS.450.1777K,10.1093/mnras/stz2495,2021MNRAS.505.3016N,2021ApJ...906...94Z,10.1093/mnras/stad606} estimate a limit on the maximum absolute magnitude of $M\geq$-16.5. 
Combining this absolute magnitude limit with the estimated decline rates and the magnitude limits on observations imposed by our follow-up strategy, we determined a maximum luminosity distance of $D\leq\dlim$\,Mpc (corresponding to a redshift limit of $z\leq\zlim$) would be feasible for both detecting kilonova candidates and allowing follow up with deeper observations to constrain the fading rate.

Transients that meet our projected host galaxy separation and distance criteria are selected automatically within our relational database in what we internally call the ``Fast Track'' list, and all are manually vetted. 
Our software automatically checks the IAU Transient Name Server (TNS)\footnote{Transient Name Server: \url{https://www.wis-tns.org/}.} for any previous discovery (and also registers our first detection, whether or not we were the first to discover it) and any submitted spectroscopic classifications. 
We then produce forced photometry at the position of the Pan-STARRS transient to check for any subthreshold detections in the Pan-STARRS data that may indicate a rising transient and place limits on the explosion epoch and rise time. 

For transients with an absolute magnitude $M\geq$-16.5 that had not been previously reported by other sky surveys and did not show evidence of a slow rise in the early forced photometry (or through other available TNS information), we triggered a follow-up epoch with Pan-STARRS on the next available night. 
Follow-up was conducted using the same filter as the discovery epoch: a single 200s exposure if the discovery magnitude was $m\geq$20.5, or 2$\times$100s exposures otherwise. 
The resulting forced photometry light curve was fed into our custom algorithm for predicting kilonova likelihood (see Section\,\ref{subsec:kilo-predict} for details). 
If the transient scored low on likelihood, a single follow-up epoch in the same filter was queued approximately 7-14 days later to quantify the slow light curve evolution. 
If it scored high, we further checked the ZTF \citep{2019PASP..131a8002B} data stream in Lasair \citep{2019RNAAS...3...26S}\footnote{https://lasair-ztf.lsst.ac.uk}, and requested ZTF and ATLAS forced photometry through the ZTF forced photometry service \citep{Masci_2019} and the ATLAS forced photometry server \citep{2018PASP..130f4505T,2020PASP..132h5002S,2021TNSAN...7....1S} across all time at the transient's position. 
If there was no evidence of historical activity in either telescope system, we assigned a Pan-STARRS follow-up campaign for the transient (nightly \grizyps\,exposures), and alerted the community to the potential kilonova candidate through the TNS AstroNote system \citep{2019TNSAN...1....1G}.

\subsection{Kilonova Prediction Algorithm} 
\label{subsec:kilo-predict}
To assist in our search, we developed an algorithm to analyse the kilonova likeness of an input light curve from transients in the ATLAS and Pan-STARRS sky surveys. 
Predictions are made by comparing the input to the parameter space of synthetic kilonova light curves (see below for details), in addition to real, well-sampled light curves collected by sky surveys over the last decade for various classes of fast-evolving transients that, when sampled sparsely in real-time, are contaminants in a kilonova search. 
Such contaminants include the rapidly fading and hydrogen-deficient ``ultra-stripped'' type Ibc supernovae (USSNe), the shock breakout component of type IIb supernovae (SB SN~IIb), the faint population of 02cx-like type Ia supernovae (SNe~Iax), outbursts from luminous blue variables (LBVs), intermediate luminosity red transients (ILRTs), luminous red novae (LRNe) and classical novae.  

The kilonova models were generated by \citet{2021MNRAS.505.3016N} using the \texttt{MOSFiT} package \citep{Guillochon_2018}. 
These are semi-analytic models based on the standard formalism given by \citet{Arnett1982}. 
The ejecta consists of three components with separate masses, velocities and opacities, corresponding to tidal, shocked and wind-driven ejecta.  
We also include shock heating of the polar ejecta by any jet launched during the merger following \cite{Piro2018}. 
The masses and velocities of each component are determined from the initial binary masses using scaling relations. 
We modelled for two fixed opacity values of 0.5\,cm$^{2}$g$^{-1}$ and 1\,cm$^{2}$g$^{-1}$, and 10\,cm$^{2}$g$^{-1}$ and 25\,cm$^{2}$g$^{-1}$, for the tidal and shocked components, respectively. 
The opacity of the wind-driven component was allowed to vary between 1.55$-$5.52\,cm$^{2}$g$^{-1}$ across all models. 
We uniformly sampled chirp masses between 0.7$-$2.0\,M$_\odot$, mass ratios between 0.4$-$1.0, disk ejecta mass fractions between 0.1$-$0.4, and viewing angles fixed at 0$\degree$, 30$\degree$, 60$\degree$, 75$\degree$ and 90$\degree$. 
For a subset of kilonova models with mass ratios $=$1, disk ejecta mass fractions $=$0.2 and default opacities, we included emission from the shock-heated cocoon in the ejecta energy with fixed opening angles of 10$\degree$, 20$\degree$ and 30$\degree$. 
Full details can be found in \citet{2021MNRAS.505.3016N}. 
In total, 17,045 kilonova models were generated. 
We caution that while this approach is likely not representative of the distributions of these parameters in nature, it does ensure that we comprehensively cover the diverse parameter space of plausible kilonovae. 
Covering a wider range of parameters can lead to more false positives in a Kilonova search; however, it will ensure that the end sample is more robust. 
A robust sample is crucial, considering that the known parameter space for kilonovae is still not firmly established. 

The algorithm begins by ingesting the template library of kilonova models and contaminants and separating their multi-band light curves into individual filters. 
These single-band light curves are then smoothed across a uniform time series using empirical polynomial modelling, curve fitting, and regression analysis of the original data.
Features such as the predicted rise time from explosion to peak brightness, peak absolute magnitude, incline rate to peak, and decline rates $\Delta$2d, $\Delta$5d, and $\Delta$10d after peak are extracted where available from the smoothed light curves thereafter. 
These features are collated per filter per transient class and converted into probability density functions (which we will refer to as ``$M-\dot{M}$ planes'') of
\begin{itemize}
    \item Peak absolute magnitude versus Predicted rise time.
    \item Peak absolute magnitude versus Incline rate.
    \item Peak absolute magnitude versus Decline rate.
    \item Colour versus Decline rate.
\end{itemize}
using a Gaussian kernel density estimation.
The $M-\dot{M}$ planes produced for each transient class are finally overlaid to produce filter-dependent $M-\dot{M}$ locus plots. 

\begin{figure}
    \centering
    \includegraphics[width = 0.99\columnwidth]{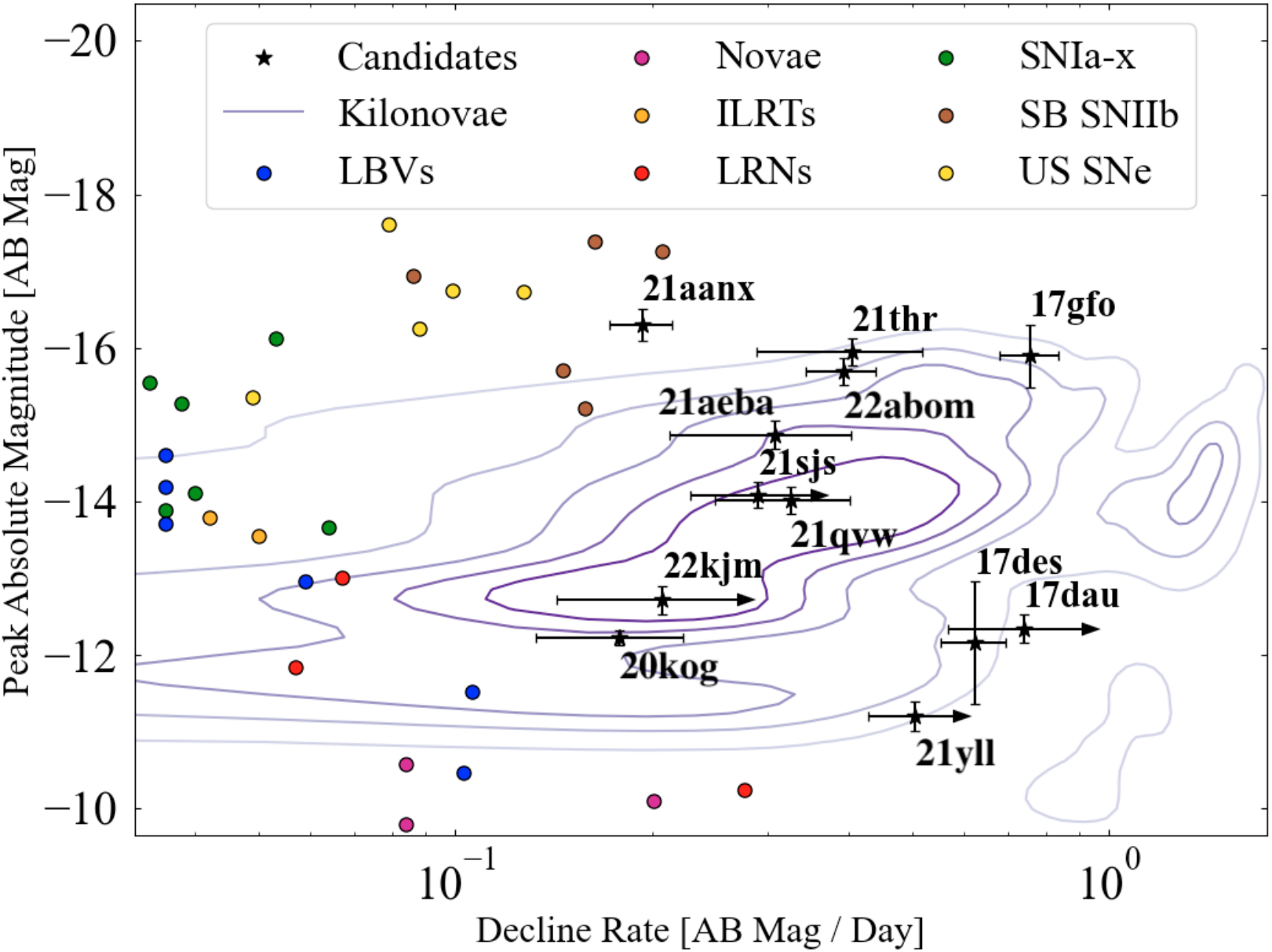}
    \caption{$M-\dot{M}$ locus plot of peak brightness versus decline rate in the \wps\,filter for the 11 fast-evolving transients identified in the Pan-STARRS sample. Our full kilonovae model distribution has been converted into a probability density function using a Gaussian kernel density estimation. It is represented by contours, where the area enclosed by darker lines depicts a higher concentration of models. Overlaid is known kilonova AT~2017gfo (17gfo) for reference.}
    \label{fig:fastfinder}
\end{figure}

Figure\,\ref{fig:fastfinder} is an example of the $M-\dot{M}$ locus plot produced for ``Peak absolute magnitude versus Decline rate'' in the \wps-filter for decline rates measured $2-5$ days after peak brightness. 
For illustrative purposes, we display only the $M-\dot{M}$ plane for the kilonovae class as contour lines, where the enclosed areas represent 60\%, 70\%, 80\%, 90\% and 98\% of the total template population, from dark-to-light shades, respectively. 
The other transient classes are presented as definitive data points to demonstrate their distribution amongst the kilonova models. 
However, $M-\dot{M}$ planes are still generated behind the scenes for these.

Once the template library has been processed, the candidate transient is prepared similarly by splitting the input light curve into the available filters.
Features are derived and projected onto the appropriate $M-\dot{M}$ locus plots. 
A probability score measuring the likeness of the input to the templates is generated for each transient class based on its position within the $M-\dot{M}$ planes. 
These probability scores are averaged across all filters involved and normalised into a single ``likelihood score'' ranging between 0\% and 100\% for each $M-\dot{M}$ locus plot.
In this kilonova search, an input that scored a kilonova resemblance of 50\% or higher in any $M-\dot{M}$ locus plot was considered a viable kilonova candidate.

We note that there are many papers in the literature proposing more sophisticated machine learning based approaches for classifying kilonova candidates from their lightcurves \citep[e.g.][]{2018ApJS..236....9N,2019PASP..131k8002M,2023A&A...677A..77B}. 
However, we have used a more straightforward approach since our search is real-time and rapid, meaning we would typically have only two epochs of imaging, and we are searching for a rapid fade combined with a faint absolute magnitude.

\section{Results of the search}
\label{sec:results}
\begin{figure}
    \centering
    \includegraphics[width = 0.99\columnwidth]{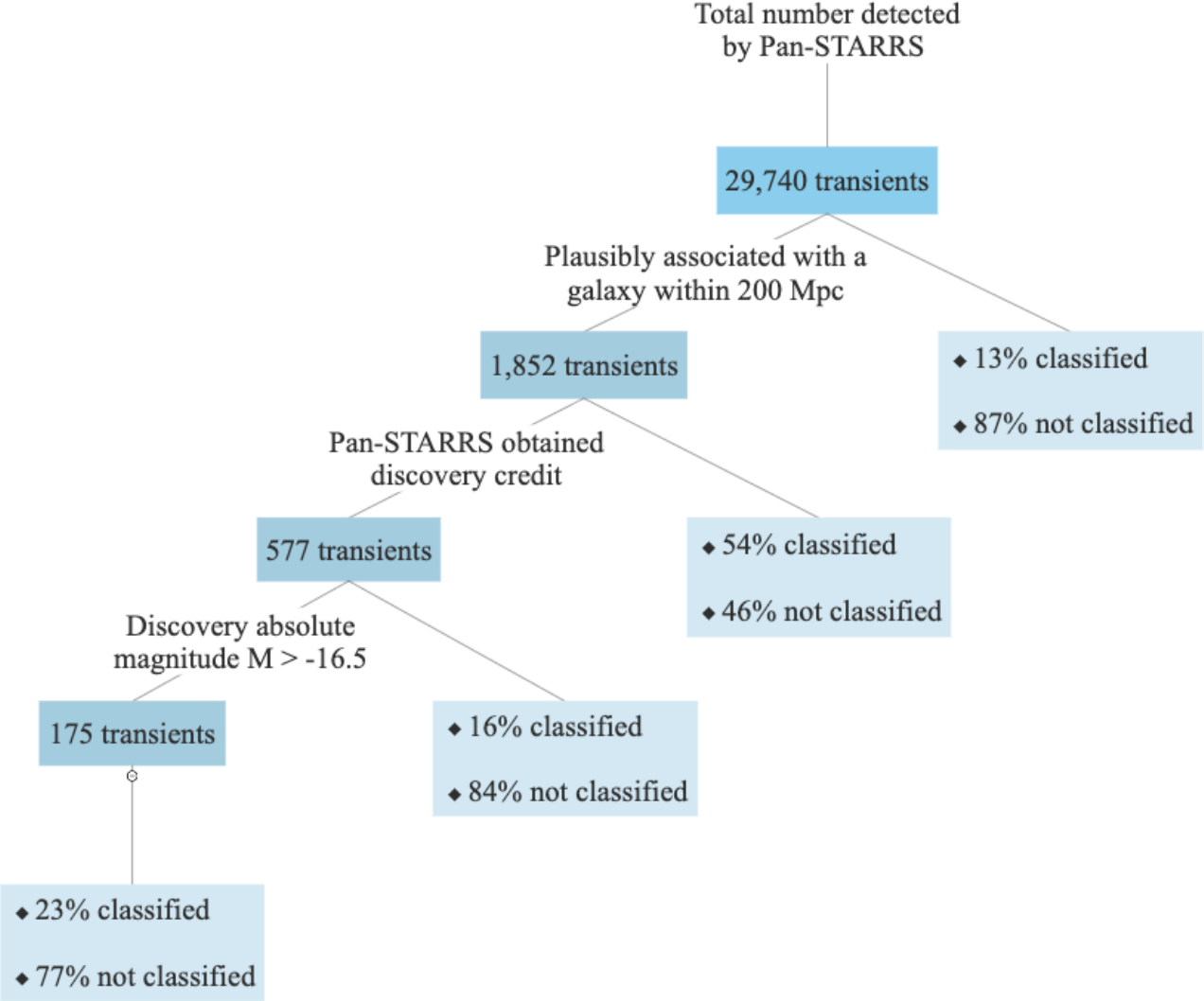}
    \caption{Tree diagram depicting the breakdown of transients detected by Pan-STARRS throughout the Pan-STARRS Search for Kilonovae. Classification status refers to spectroscopic classification only.}
    \label{fig:treeDiagram}
\end{figure}

Over 3.14 years (26th October 2019 - 15th December 2022), Pan-STARRS had detected a total of 29,740 extragalactic transient events, whose detections were reported to the TNS. 
Of these, 1,852 were plausibly associated with a galaxy within \dlim\,Mpc and subsequently appeared within our internal ``Fast Track'' list. 
Hence, around 1.6 transients per day are detected by Pan-STARRS (but not necessarily discovered first) within \dlim\,Mpc. 
We note that the redshift completeness of available galaxy catalogues will limit this. 
Pan-STARRS received discovery credit for approximately one-third of these nearby transient events, of which 175 had a discovery absolute magnitude $M>-16.5$, meeting our criteria for additional follow-up. 
This breakdown is also depicted in Figure\,\ref{fig:treeDiagram}.

Of these 175 faint transients, two were previously reported to TNS by PS1 in 2017 but exhibited outbursts in the aforementioned time frame, which PS2 independently detected. 
Approximately 75\% of this faint population was found to reside in spiral galaxies. 
This is to be expected given that the galaxy composition of the local universe is 72\% spiral \citep{Delgado_Serrano_2010} and the progenitor systems for many extragalactic transients involve young and massive stars. 

\begin{figure*}
    \centering
    \includegraphics[width = 0.99\textwidth]{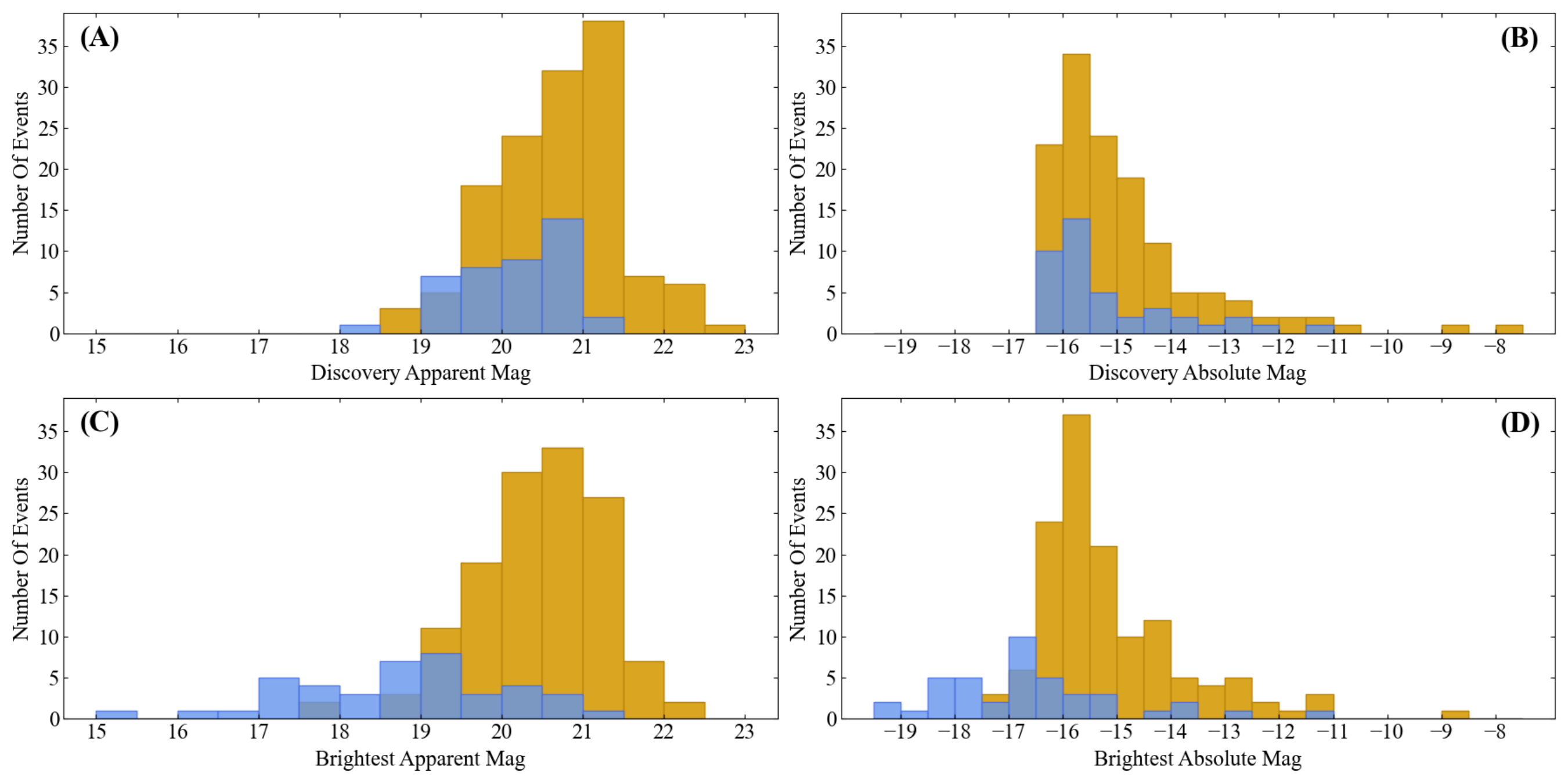}
    \includegraphics[width = 0.99\textwidth]{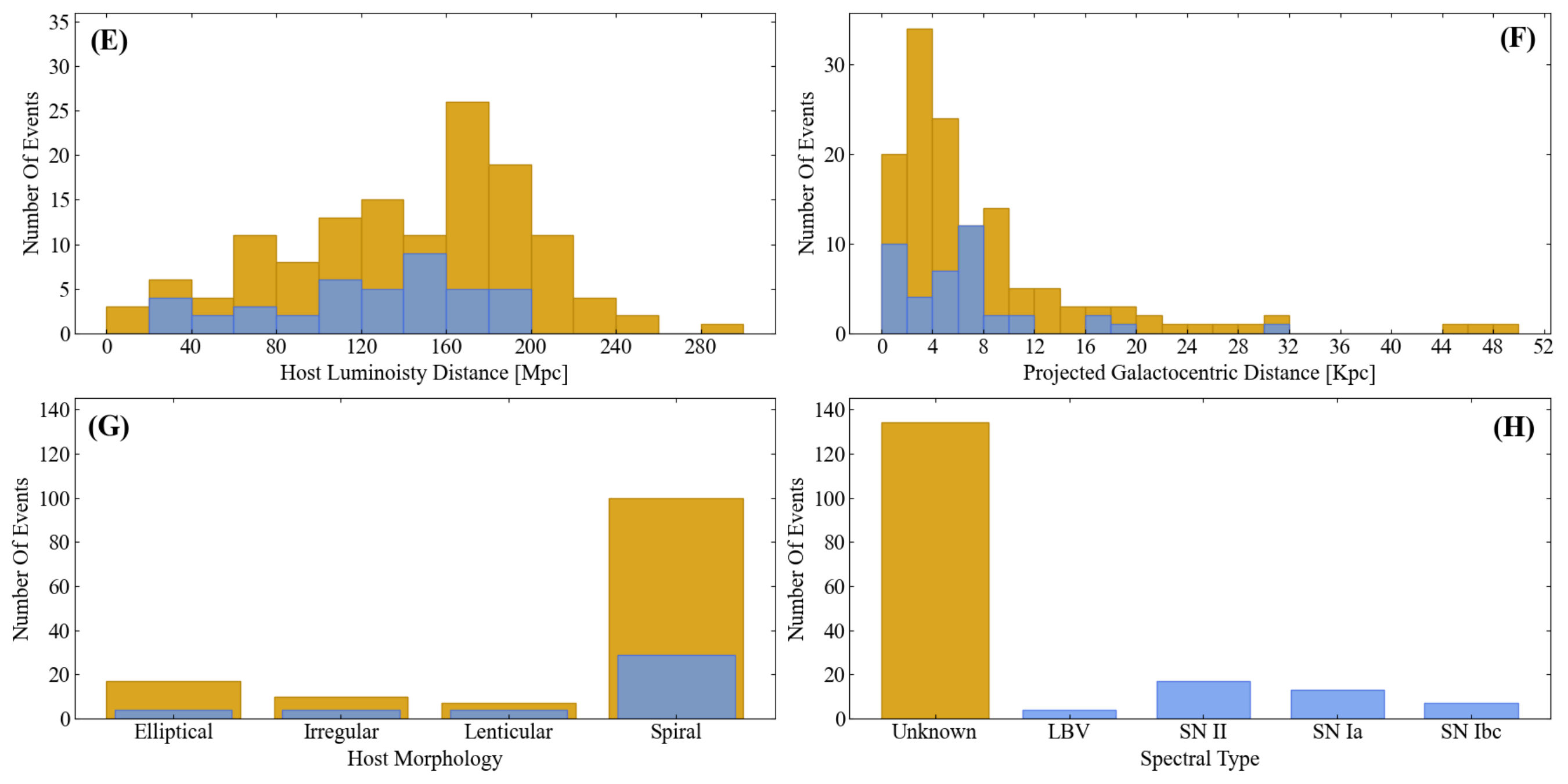}
    \caption{The discoveries of transients, mostly supernovae, are summarised in these plots: (\textbf{A}) histogram of discovery apparent mag. (\textbf{B}) histogram of discovery absolute mag. (\textbf{C}) histogram of brightest apparent mag. (\textbf{D}) histogram of brightest absolute mag. (\textbf{E}) histogram of distances using $H_{0}$. (\textbf{F}) histogram of projected offsets from host galaxy centres. (\textbf{G}) bar chart of host morphologies. (\textbf{H}) bar chart of spectral classifications. The colour of the bars refers to the transients with a spectral classification (blue) and those without (orange).}
    \label{fig:SummaryHistograms}
\end{figure*}

A statistical summary of this population is presented in Figure\,\ref{fig:SummaryHistograms}, where transients with spectroscopic classification are represented by blue bars and those without by orange bars. 

The brightest magnitude measurements represent those measured by Pan-STARRS only and do not consider any measurements made by other survey groups. 
For a small subset of transients, a significant discrepancy exists in their respective host galaxy $D_{\rm d}$ and $D_{\rm z}$ quoted distances on NED. 
In such cases, the $D_{\rm d}$ values obtained from methods such as the colour-magnitude relation \citep{1978ApJ...223..707S}, the size–magnitude relation \citep{1986AJ.....91...76R}, the Tully–Fisher relation \citep{1977A&A....54..661T,TullyFisher2013}, and the Fundamental Plane \citep{1987ApJ...313...59D,1987ApJ...313...42D} suggest smaller distances, which are consequently favoured by our Sherlock algorithm. 
To maintain consistency, all absolute magnitudes, projected galactocentric distances, and host luminosity distances presented in Figure\,\ref{fig:SummaryHistograms} have been determined through the host redshift using  $H_{0}=$\HO$\pm5$\kms; however, our analyses of the kilonova candidates in Section\,\ref{subsec:fast11} use whichever distance is smaller. 
Host morphologies, if available, were obtained from NED and the SIMBAD Astronomical Database \citep{refId0} and split into four general types: spiral, elliptical, lenticular and irregular. 
Those that did not have a predetermined morphology were assigned one manually by comparing their visual appearance in the Pan-STARRS 3$\pi$ survey images \citep{2016arXiv161205560C} to morphology templates listed in \cite{Kennicutt_2003} and \cite{2010ApJS..186..427N}. 
The spectral classifications of the transients represent those reported on TNS.

Only 23\% of this faint transient population received a spectroscopic classification. 
This is presumably due to the fact that the majority (80\%) of these transients had a brightest apparent magnitude $m$>20. 
No routine 2$-$4\,m telescope classification programmes that systematically target such faint transients are running. 
For example, the PESSTO and ePESSTO+ surveys employ a limit of $m$<19.5 \citep{2015A&A...579A..40S}, the SCAT survey employs a limit of $m$<19 \citep{Tucker_2022} and the ZTF bright transient survey employs a limit of $m$<18.5 \citep{Fremling_2020,Perley_2020} for their classifications, respectively. 
Our target magnitude is, however, comfortably reached with 6$-$10\,m class telescopes, but such large facilities are not currently employed for systematic classification surveys. 

\subsection{Kilonova Candidates and Contaminants}
\label{subsec:fast11}
Eight of the 175 faint transients were flagged by our algorithm as plausible kilonova candidates in real-time. 
Another three displayed fast-evolving outbursts during the search, which in isolation would resemble kilonovae, but with earlier detections indicating an underlying variable source rather than a one-off transient. 
This includes the two transients first discovered in 2017, as previously mentioned. 

\begin{figure}
    \centering
    \includegraphics[width = 0.95\columnwidth]{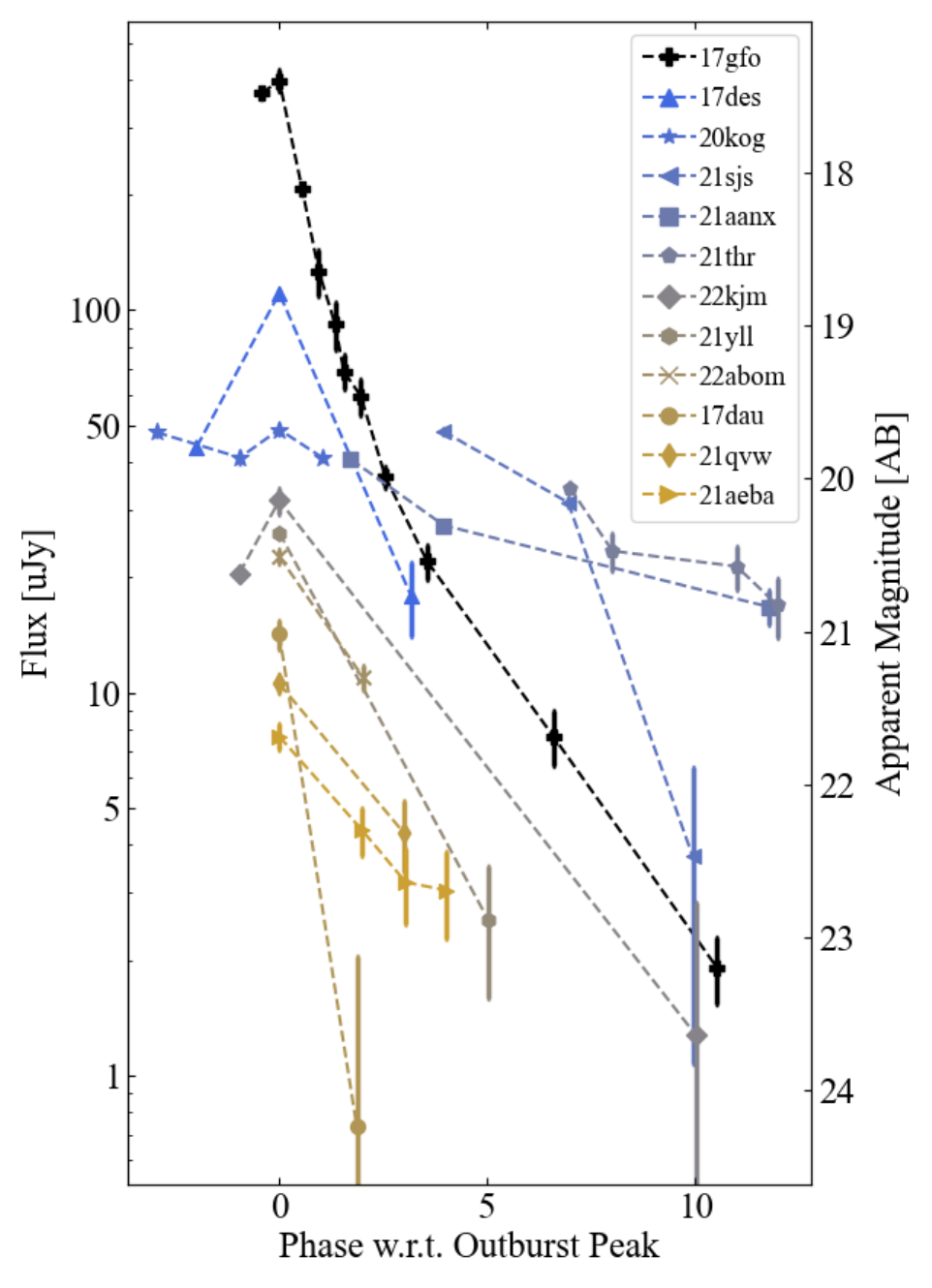}
    \caption{The \wps\,filter light curves of the fastest outburst measured for the 11 fast-evolving transients. Phase is with respect to the outburst peak measured from the combined Pan-STARRS and ATLAS forced photometry if ATLAS data were available. The constructed \wps\,filter light curve of known kilonova AT~2017gfo (17gfo) has been included for comparison.}
    \label{fig:outburstlcs}
\end{figure}

An $M-\dot{M}$ locus plot containing all 11 identified transients is shown in Figure\,\ref{fig:fastfinder}, where the decline rate and absolute magnitude represent the fade measured 2$-$5 days after peak brightness in the \wps\,filter. 
For transients with more than one outburst recorded, we present their fastest outburst in the figure.
When calculating decline rates between detections and non-detections, a 3$\sigma$ magnitude limit is assumed for the latter and denoted by arrows. 
The light curves pertaining to the fast evolution demonstrated by these transients are shown in Figure\,\ref{fig:outburstlcs}, and their physical properties are summarised in Table\,\ref{tab:fastobjs}. 
Below, we discuss each transient in turn and assess its plausibility as a kilonova candidate. 
All magnitudes and times quoted represent the nightly stacked forced photometry fluxes. 
All absolute magnitudes quoted have been corrected for foreground Galactic dust extinction but have not been corrected for any host galaxy extinction.

\begin{figure*}
    \centering
    \includegraphics[width = 0.90\textwidth]{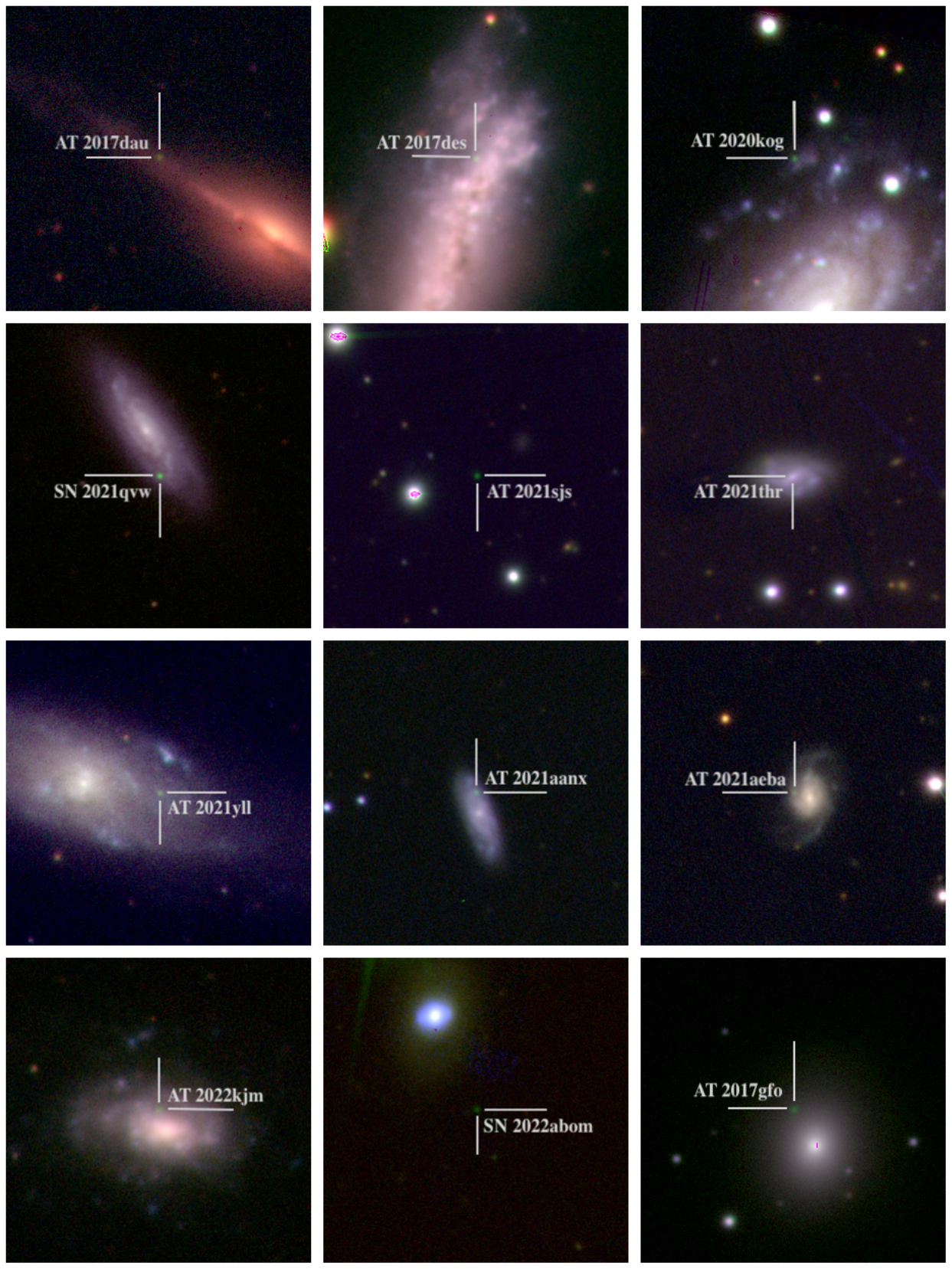}
    \caption{A mosaic of the 11 fast-evolving transients within their hosts. These GRB colour composites were created using the \wps\,target image obtained for each transient at their brightest epoch and \gps\,and \ips\,reference images from the Pan-STARRS 3$\pi$ survey. Each composite is centred on its respective transient (green dot) and measures $75\arcsec\times75\arcsec$ wide. The majority of our contaminants were found in star-forming hosts. A colour composite of known kilonova AT~2017gfo alongside its host has been included for reference.}
    \label{fig:mosaic}
\end{figure*}

The location of the transients within their host galaxies is presented in Figure\,\ref{fig:mosaic} along with the host of AT~2017gfo to illustrate the differences between the systems of our faint and fast-evolving candidates and a true kilonova. 

\begin{table*}
\centering
\begin{tabular}{ccccccccc}
\hline
Internal & IAU      & RA          & Dec          & Absolute Mag        & Decline Rate         & Host Distance    & Galactocentric & Not kilonova \\
Name     & Name     & hh:mm:ss.ss & dd:mm:ss.ss  & ($M_{\rm \wps}$) & ($mag d^{-1}$) & (Mpc) & Distance (Kpc) & Reason \\
\hline   
PS17chm  & AT~2017dau  & 14:14:17.09 & +35:25:43.64 & $-12.35 \pm 0.19$   & $0.74 \pm 0.18$      & $46\pm4$         & $7.4$          & LBV    \\
PS17cke  & AT~2017des  & 12:34:18.89 & +06:28:26.94 & $-12.20 \pm 0.80$   & $0.62 \pm 0.07$      & $15\pm5$         & $2.4$          & Multiple bursts   \\
PS20dgq  & AT~2020kog  & 16:18:47.65 & +07:25:16.86 & $-12.23 \pm 0.09$   & $0.18 \pm 0.05$      & $22\pm1$         & $4.2$          & Pre-rise activity \\
PS21gqd  & SN~2021qvw  & 23:16:38.29 & +15:53:40.01 & $-14.02 \pm 0.17$   & $0.33 \pm 0.08$      & $111\pm8$        & $6.1$          & LBV / SN~IIn  \\
PS21hlm  & AT~2021sjs  & 20:33:22.78 & -07:59:46.62 & $-14.09 \pm 0.17$   & $0.31 \pm 0.05$      & $54\pm4$         & $44.8$         & Likely CV  \\
PS21hzg  & AT~2021thr  & 16:00:12.73 & +16:24:41.29 & $-15.95 \pm 0.17$   & $0.40 \pm 0.12$      & $153\pm11$       & $0.4$          & Long-lived     \\
PS21juu  & AT~2021yll  & 01:41:05.87 & -05:34:07.67 & $-11.21 \pm 0.19$   & $0.48 \pm 0.08$      & $20\pm2$         & $1.9$          & Multiple bursts  \\
PS21ksy  & AT~2021aanx & 23:19:18.88 & +18:36:53.20 & $-16.30 \pm 0.21$   & $0.19 \pm 0.02$      & $164\pm17$       & $3.3$          & Long-lived     \\
PS21mbb  & AT~2021aeba & 08:08:45.18 & +18:42:06.11 & $-14.87 \pm 0.19$   & $0.31 \pm 0.06$      & $195\pm14$       & $3.7$          & Multiple bursts \\
PS22ehy  & AT~2022kjm  & 13:06:29.05 & -13:34:12.38 & $-12.72 \pm 0.19$   & $0.21 \pm 0.06$      & $36\pm3$         & $1.0$          & Multiple bursts  \\
PS22luo  & SN~2022abom & 03:41:09.14 & -02:50:54.85 & $-15.69 \pm 0.18$   & $0.39 \pm 0.05$      & $158\pm11$       & $18.2$         & SN~Ia early excess \\
\hline
\end{tabular}
\caption{A list of the physical properties for the 11 fast-evolving transients identified in the Pan-STARRS search for Kilonovae. Their reason for elimination as a kilonova candidate is explained thoroughly within the text. The absolute magnitudes and decline rates are measured in the \wps\,filter.}
\label{tab:fastobjs}
\end{table*}

\subsubsection*{\textbf{AT~2017dau}}
PS1 discovered AT~2017dau on MJD 57855.482 (2017-04-12 11:34:04 UTC), observing the transient on a single night at $m_{\rm \ips}=19.31\pm0.04$. 
AT~2017dau was found in the outskirts of the edge-on and dusty lenticular/spiral host galaxy UGC 09113, which has a luminosity distance of $D_{\rm L}=46\pm4$\,Mpc, equivalent to a distance modulus of $\mu=33.31\pm0.17$, implying an absolute magnitude of $M_{\rm \ips}=-14.04\pm0.17$ at the time of discovery. 
AT~2017dau was classified ten days later as an LBV outburst owing to the strong and narrow H${\rm \alpha}$ and H${\rm \beta}$ P-cygni profiles imposed on a black body continuum of $T_{\rm eff}\simeq5500$\,K \citep{2017TNSCR.463....1B}. 
Over the following six years, Pan-STARRS observed many smaller amplitude outbursts at the position of AT~2017dau, reminiscent of S Doradus (S Dor) variability \citep{humphreys1994luminous}. 
On MJD 59641.627 (2022-03-03 15:02:46 UTC), another eruption of AT~2017dau was observed by PS2 with a peak absolute magnitude of $M_{\rm \wps}=-12.35\pm0.19$ and a very rapid decline rate of $dm_{\wps}/dt=0.74\pm0.18$, fading from view within two days. 
While we securely ruled out AT~2017dau as a kilonova due to its previous activity, the remarkably rapid decline of the 2022 outburst, as illustrated in Figure\,\ref{fig:fastfinder} and Figure\,\ref{fig:outburstlcs}, shows that LBV outbursts are potential contaminating sources in kilonova searches. 

\subsubsection*{\textbf{AT~2017des}}
AT~2017des was originally flagged as a potential kilonova candidate in a previous search of Pan-STARRS data conducted by \cite{2021MNRAS.500.4213M}. 
PS1 discovered AT~2017des on MJD 57859.308 (2017-04-16 07:24:06 UTC) at $m_{\rm \wps}=19.69\pm0.04$ in the irregular host galaxy NGC 4532, which has a luminosity distance of $D_{\rm L}=15\pm5$\,Mpc ($\mu=30.9\pm0.8$) favoured by \cite{2021MNRAS.500.4213M}, implying an absolute magnitude of $M_{\rm \wps}=-11.2\pm0.8$ at the time of discovery. 
Four attempts were made to obtain a spectrum of AT~2017des; unfortunately, the transient lies within a high surface brightness region of the host, which made target acquisition of the rapidly fading source difficult. 
No obvious transient in the spectroscopic 2D images could be identified in all cases. 
However, during this search for kilonovae with Pan-STARRS,  PS2 observed another rapid outburst at the position of AT~2017des on MJD 59618.493 (2022-02-08 11:49:56 UTC) with a peak absolute magnitude of $M_{\rm \wps}=-12.2\pm0.8$ and decline rate of $dm_{\rm \wps}/dt=0.62\pm0.07$ \citep{2022TNSAN..36....1F}.
\cite{2021MNRAS.500.4213M} had proposed an LBV outburst was a likely explanation for the fast transient AT~2017des, and these successive outbursts at the same position confirm the likely LBV nature. 
However, again, the data illustrate that such outbursts can be kilonova contaminants.

\subsubsection*{\textbf{AT~2020kog}}
We flagged AT~2020kog as an intrinsically faint transient source with a peak apparent magnitude of $m_{\rm \wps}=19.68\pm0.03$ on MJD 58993.387 (2020-05-24 09:17:53.952 UTC). 
AT~2020kog was found in the spiral host galaxy NGC 6106, which has a mean direct distance of $D_{\rm L}=22\pm1$\,Mpc ($\mu=31.75\pm0.09$; from NED), implying a peak absolute magnitude of $M_{\rm \wps}=-12.23\pm0.09$. 
AT~2020kog displayed a short oscillation at peak brightness with a moderate fade of $dm_{\rm \wps}/dt=0.18\pm0.05$. 

\begin{figure}
    \centering
    \includegraphics[width = 0.95\columnwidth]{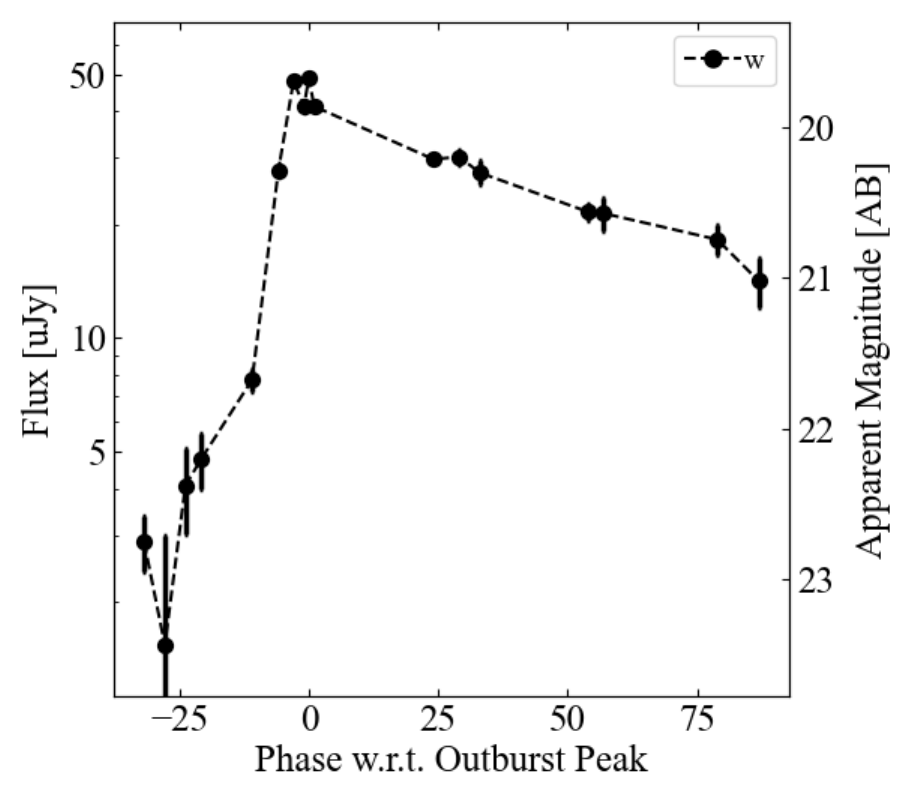}
    \includegraphics[width = 0.95\columnwidth]{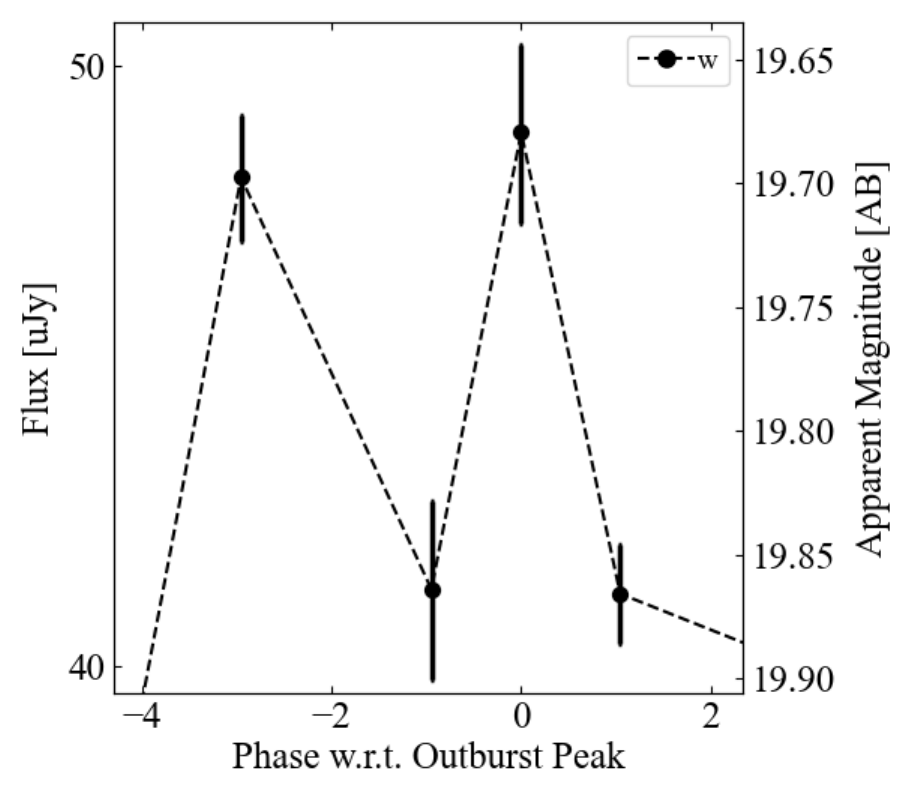}
    \caption{Top panel: The full \wps filter light curve of AT~2020kog, displaying the pre-rise activity before the quick outburst and the slow declining tail thereafter. Bottom panel: Enhanced view of the peak brightness of AT~2020kog demonstrating the short and rapid fluctuating behaviour.}
    \label{fig:2020koglc}
\end{figure}

This drew our attention as a potential faint and fast-fading transient. 
However a lower significance excess was detected 32 days before the peak luminosity on MJD 58961.551 (2020-04-22 12:41:46 UTC) at $m_{\rm \wps}=22.75\pm0.16$ and forced photometry confirmed the existence of pre-rise activity, quickly ruling the transient out as a kilonova candidate. 
The full \wps\,filter light curve of AT~2020kog and its relatively rapid oscillation at peak brightness are shown in Figure\,\ref{fig:2020koglc}, illustrating a slowly decaying tail lasting approximately 90 days observed through normal survey operations. 
Nonetheless, the behaviour exhibited by AT~2020kog at peak brightness was unusual, and we released an AstroNote highlighting the behaviour at the time \citep{2020TNSAN.107....1F}.

\subsubsection*{\textbf{SN~2021qvw}}
\begin{figure}
    \centering
    \includegraphics[width = 0.95\columnwidth]{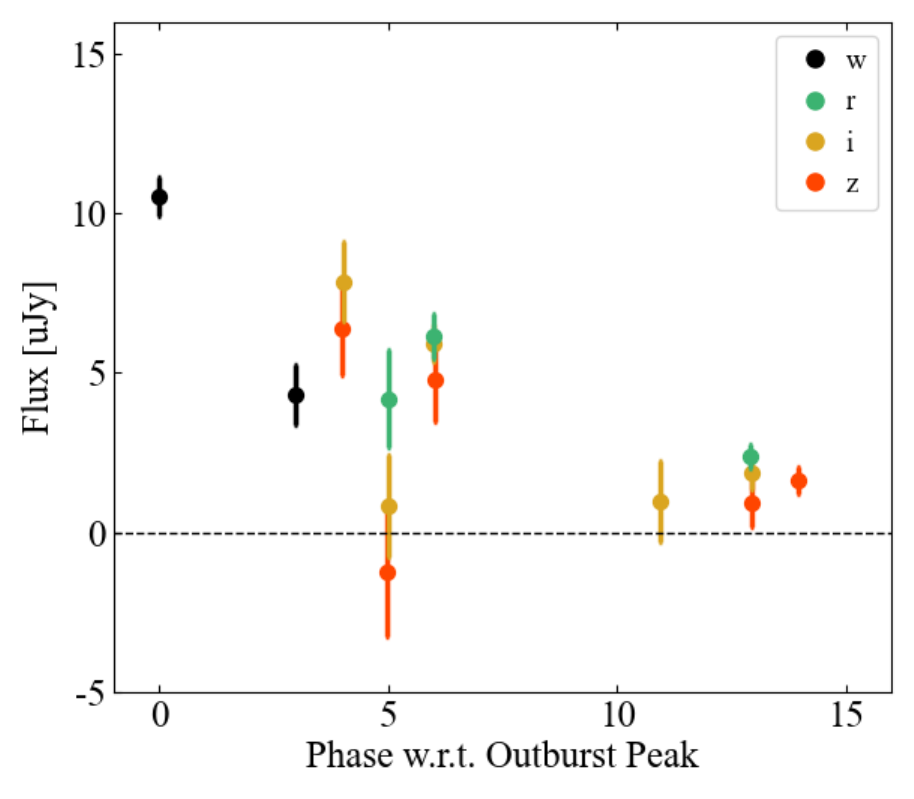}
    \includegraphics[width = 0.95\columnwidth]{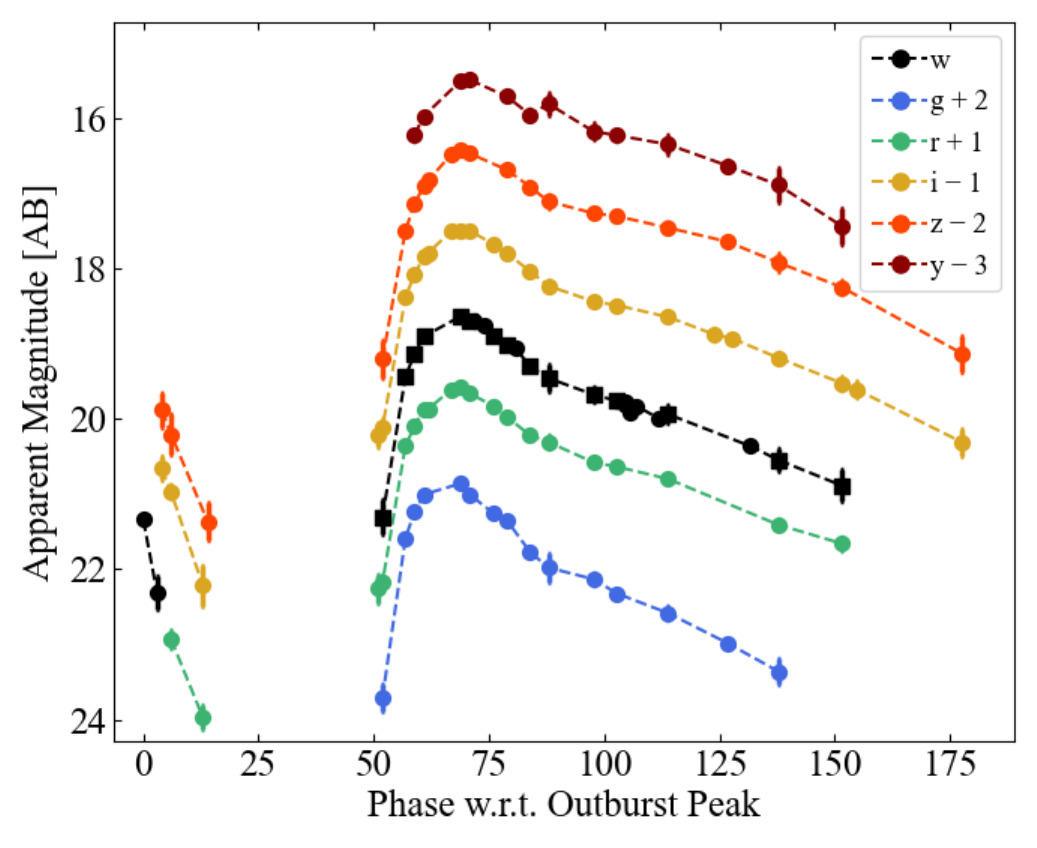}
    \caption{Top panel: Pan-STARRS photometry of the first 14 days of the SN~2021qvw outburst. Bottom panel: The full Pan-STARRS 3$\sigma$ light curve of SN~2021qvw displaying the initial outburst and the later SN explosion. Square markers on the \wps\,filter represent synthetic values derived from the \grips\,filter measurements made on the same night.}
    \label{fig:2021qvwlc}
\end{figure}

Pan-STARRS discovered SN~2021qvw on MJD 59386.570 (2021-06-21 13:40:59 UTC) at $m_{\rm \wps}=21.34\pm0.06$ in the spiral host galaxy CGCG 454-019 which has a luminosity distance of $D_{\rm L}=111\pm8$\,Mpc ($\mu=35.22\pm0.16$), implying an absolute magnitude of $M_{\rm \wps}=-14.02\pm0.17$. 
Follow-up of SN~2021qvw three days later measured $m_{\rm \wps}=22.31\pm0.21$ and a decline rate of $dm_{\rm \wps}/dt=0.33\pm0.08$, placing it near the centre of the kilonova probability density within Figure\,\ref{fig:fastfinder}. 
A check of the ZTF and ATLAS forced photometry at the location of SN~2021qvw revealed no historical activity, and so an AstroNote was released highlighting the transient as a potential kilonova candidate \citep{2021TNSAN.179....1F}. 
Pan-STARRS followed SN~2021qvw extensively over the following 14 days, collecting exposures in the \rizps\,filters as shown in the top panel of Figure\,\ref{fig:2021qvwlc}. 
On MJD 59392.590 (2021-06-27 14:08:53), approximately six days after discovery, the transient had brightened slightly, and we measured a colour $\rps-\zps = -0.28\pm0.29$ suggesting no significant colour evolution had occurred, unlike AT~2017gfo at similar times (see the end of Section\,\ref{sec:disc} for additional details). 
Pan-STARRS observed SN~2021qvw through normal survey operations 50 days later, on MJD 59437.41 (2021-08-11, 9:53:16 UTC) and found the transient had reached brightness levels similar to those measured at the time of discovery. 
Subsequent scheduled observations revealed the transient was rapidly rising ($>0.5$\,mag\,d$^{-1}$) in all filters. 
A spectrum was obtained eight days later ($\sim$60 days after discovery), revealing a blue continuum with strong and narrow H${\rm \alpha}$ in emission and classifying the early flux of SN~2021qvw as an LBV type of outburst that then rose to supernova luminosities \citep{2021TNSAN.221....1F,2021TNSCR2879....1L}. 
LBV outbursts as precursors to SN-like explosions have been documented in the past, for example, SN~2009ip \citep{2010AJ....139.1451S,2011ApJ...732...32F,2013MNRAS.430.1801M,2013ApJ...767....1P,2014ApJ...787..163G,2014MNRAS.438.1191S}, UGC~2773-OT \citep{2010AJ....139.1451S,2016MNRAS.455.3546S}, SN~2010mc \citep{2013Natur.494...65O,2014MNRAS.438.1191S} and LSQ13zm \citep{2016MNRAS.459.1039T}. 
The full 3$\sigma$ light curve of SN~2021qvw, as observed by Pan-STARRS, is shown in the bottom panel of Figure\,\ref{fig:2021qvwlc}.
Nevertheless, the early data of this faint and fast-fading transient source is quite similar to kilonova light curves.

\subsubsection*{\textbf{AT~2021sjs}}
\begin{figure}
    \centering
    \includegraphics[width = 0.95\columnwidth]{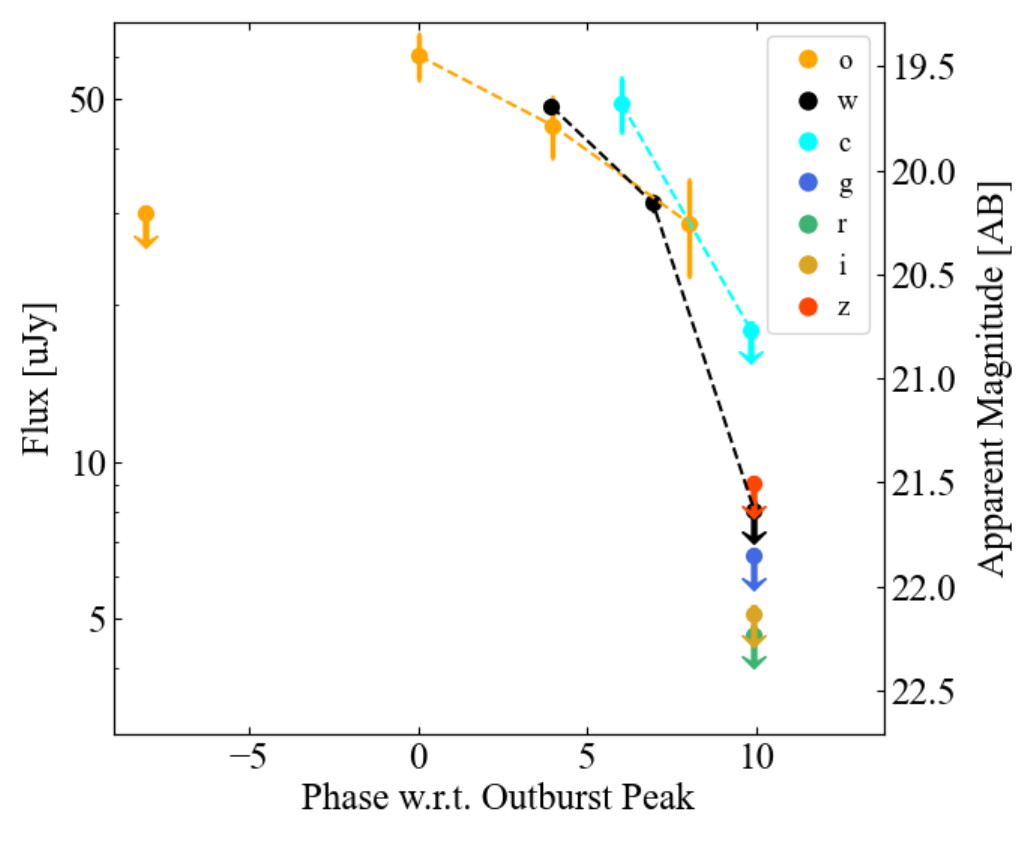}
    \includegraphics[width = 0.75\columnwidth]{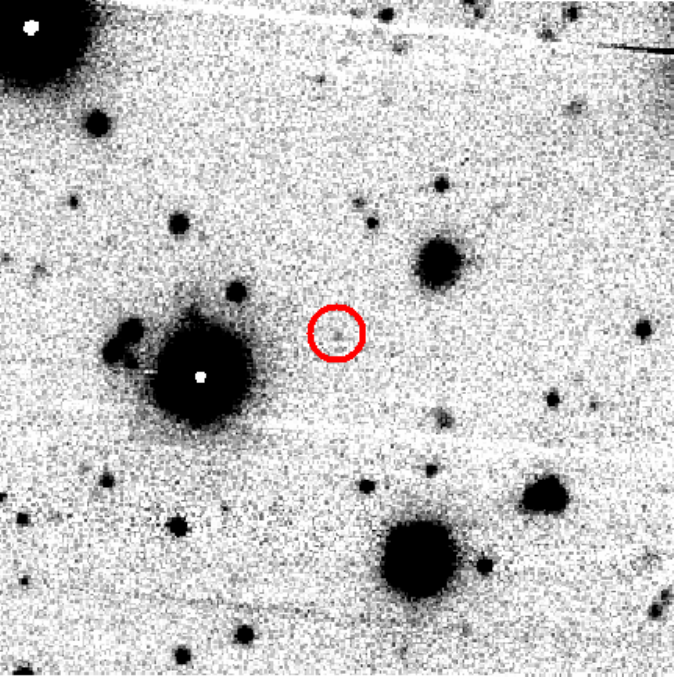}
    \caption{Top panel: The Pan-STARRS $grizw_{\rm PS}$ and ATLAS $co_{\rm A}$ filter light curve for AT~2021sjs. Data points with downward-pointing arrows represent limits at 3$\sigma$. The dashed lines highlight the fast-decaying trend. Bottom panel: A \wps\,filter reference image constructed on MJD 57122.612 (2015-04-10 14:41:15 UTC). The position of AT~2021sjs is highlighted inside the red circle and coincides with a faint point-like source.}
    \label{fig:2021sjslc}
\end{figure}

AT~2021sjs was discovered on MJD 59400.454 (2021-07-05 10:53:46 UTC) at $m_{\rm \wps}=19.69\pm0.02$ and a high projected offset of 44.8\,Kpc from the spiral galaxy 6dF J2033110-080023 which is at a luminosity distance of $D_{\rm L}=54\pm5$\,Mpc ($\mu=33.65\pm0.17$), implying an absolute magnitude of $M_{\rm \wps}=-14.09\pm0.17$. 
Follow-up three days later showed the transient had faded by 0.47 magnitudes, implying an initial decline rate of $dm_{\rm \wps}/dt=0.15\pm0.01$. 
However, additional observations performed on MJD 59406.453 (2021-07-11 10:52:19.200 UTC) revealed AT~2021sjs had faded entirely from view, proposing a considerably faster decline rate of $dm_{\rm \wps}/dt=0.31\pm0.05$. 
Pan-STARRS forced photometry showed no sign of previous activity at the transient's location, and no further detections of outbursts or excess flux have been observed by Pan-STARRS (up to March 2025). 
There was no independent 5$\sigma$ detection by ATLAS, but ATLAS forced photometry around the time of the Pan-STARRS discovery returned excess flux in both cyan and orange filters. 
The colours suggest an intrinsically blue transient at peak (see top panel of Figure\,\ref{fig:2021sjslc}). 
While no underlying source is visible in the Pan-STARRS 3$\pi$ survey images \citep{flewelling2020}, a faint point-like source is identifiable in our proprietary deep \wps\,filter NEO survey images. 
In the bottom panel of Figure\,\ref{fig:2021sjslc}, we provide the \wps\,filter reference image constructed on MJD 57122.612 (2015-04-10 14:41:14.881 UTC) through the stacking of $40\times45$\,s exposures. 
The red circle highlights the position of AT~2021sjs and coincides with this faint source, which we estimate has magnitude \rps=$24.5\pm0.2$ (assuming the \wps\,filter can be transformed to \rps\,filter with no colour correction).
The transient's large offset from galaxy 6dF J2033110-080023, coincidence with a faint apparent point source, the fast-fading nature and intrinsic blue colour at peak, all point toward a Galactic CV as the most probable explanation.
Higher resolution and deeper images of the host object would be needed to confirm its stellar nature. 

\subsubsection*{\textbf{AT~2021thr}}
\begin{figure}
    \centering
    \includegraphics[width = 0.95\columnwidth]{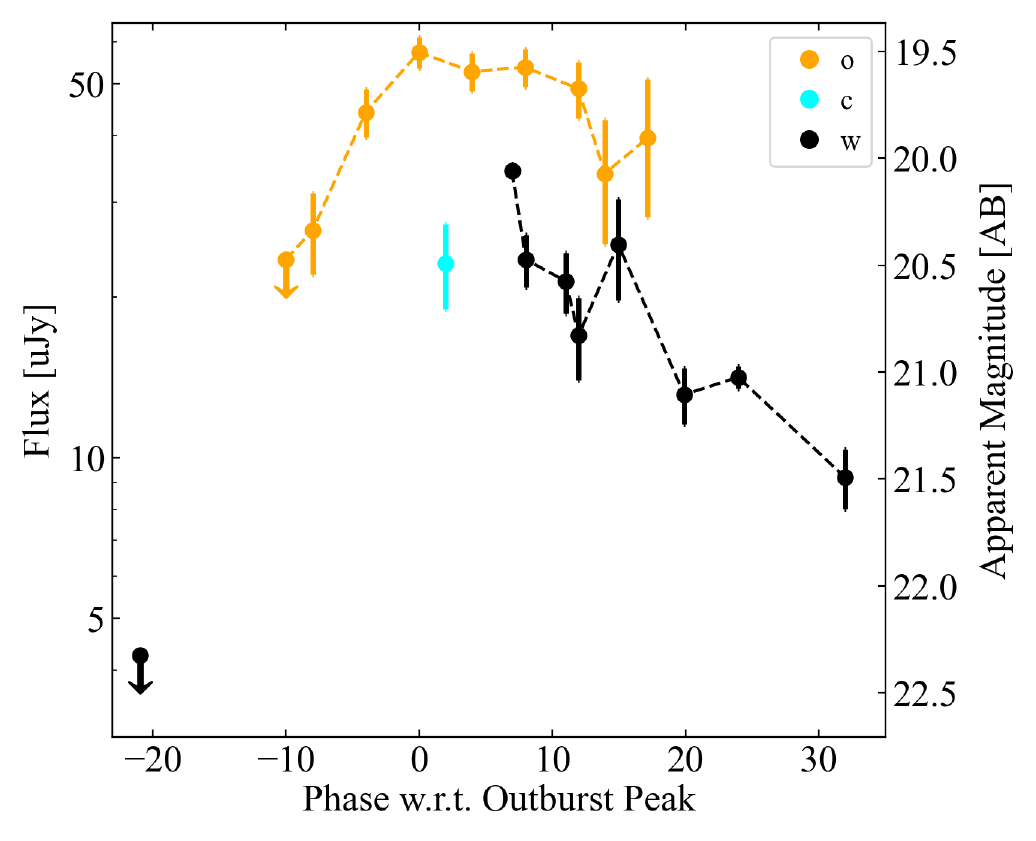}
    \caption{The full ATLAS + Pan-STARRS light curve for AT~2021thr. Data points with downward-pointing arrows represent limits at 3$\sigma$. The dashed lines highlight the possible undulations observed in both telescope systems.}
    \label{fig:2021thrlc}
\end{figure}

AT~2021thr was discovered on MJD 59408.364 (2021-07-13 08:44:10 UTC) with a apparent magnitude of $m_{\rm \wps}=20.06\pm0.03$. 
The transient's astrometric position is within 0.5\arcsec\ from the centre of the spiral host galaxy SDSS J160012.74+162440.8, which has a luminosity distance of $D_{\rm L}=153\pm11$\,Mpc ($\mu=35.93\pm0.17$) and therefore suggests an absolute magnitude of $M_{\rm \wps}=-15.95\pm0.17$. 
A  0.5\arcsec offset would correspond to a projected distance of 
0.4\,Kpc.
Typically, we would classify such transients as being coincident with the nucleus. 
The last forced photometry measurement from Pan-STARRS occurred 28 days before discovery, and we measured a $3\sigma$ limit of $m_{\rm \wps}>22.3$. 
Initial follow-up on MJD 59409.391 (2021-07-14 09:23:02 UTC) measured $m_{\rm \wps}=20.47\pm0.11$ and a fast decline rate of $dm_{\rm \wps}/dt=0.40\pm0.12$. 
Additional Pan-STARRS \wps\,filter observations over the following 25 days demonstrated an unusual light curve, with a more gradual fade overall (1.6 mag fade from discovery to +25 days) with some evidence of undulations.
We further checked ATLAS forced photometry, and although ATLAS made no independent 5$\sigma$ detection at the time, we did find excess flux rising 10$-$12 days before the Pan-STARRS discovery and a fade after peak brightness similar to that of Pan-STARRS.
When considering the full ATLAS and Pan-STARRS light curve (see Figure\,\ref{fig:2021thrlc} for reference), we measure a peak absolute magnitude of $M_{\rm \oat}=-16.51\pm0.18$ (with only foreground extinction correction) and an intrinsic red colour ($\cat-\oat=0.91\pm0.21$) at peak.
The prolonged and relatively bright nature of the light curve rules out AT~2021thr as a kilonova candidate. 
An intrinsically faint flare related to the nucleus or a highly extinguished supernova is more likely.

\subsubsection*{\textbf{AT~2021yll}}
\begin{figure}
    \centering
    \includegraphics[width = 0.95\columnwidth]{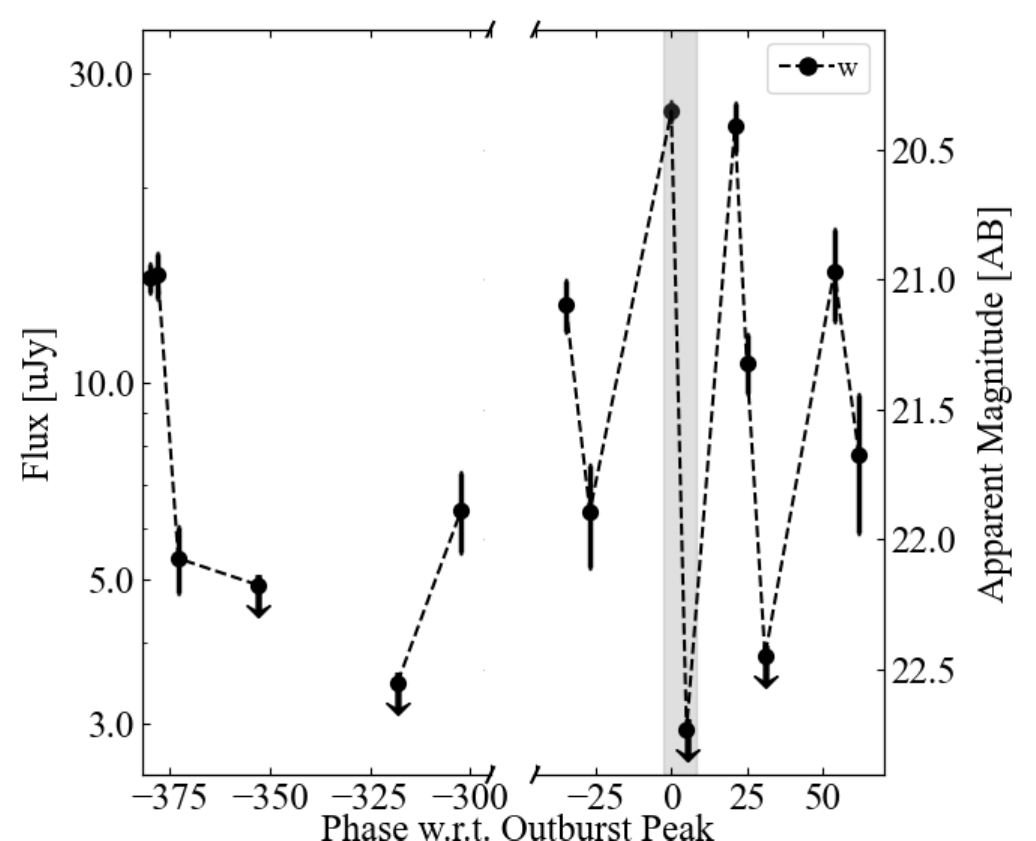}
    \caption{The Pan-STARRS \wps\,filter light curve for AT~2021yll. Data points with downward-pointing arrows represent limits at 3$\sigma$. The high amplitude, short-scale flickering occurred approximately 300\,days after the Pan-STARRS discovery. The grey-shaded region highlights the fastest and brightest outburst that was observed, which is a compelling kilonova contaminant when isolated.}
    \label{fig:2021ylllc}
\end{figure}

Pan-STARRS discovered AT~2021yll on MJD 59465.571 (2021-09-08 13:42:14 UTC) with an apparent magnitude of $m_{\rm \wps}=20.99\pm0.05$ in a dense star-forming region of the irregular host galaxy MCG -01-05-017 (alternatively classified as a type 2 Seyfert galaxy), which has a mean direct distance of $D_{\rm L}=20\pm2$\,Mpc ($\mu=31.48\pm0.19$; from NED). 
Follow-up observations conducted two days later revealed a flat light curve. 
However, subsequent observations on MJD 59472.584 (2021-09-15 14:01:04.512 UTC) showed that the transient had faded by one magnitude to $m_{\rm \wps}=22.07\pm0.12$, corresponding to a moderate decline rate of $dm_{\rm \wps}/dt=0.21\pm0.03$.
Unfortunately, poor observing conditions and weather prevented any additional follow-up.
Approximately 80 days later, another outburst was observed through normal survey operations before the transient disappeared behind the Sun.
Once the transient emerged from solar conjunction, Pan-STARRS again observed a series of outbursts at the location of AT~2021yll through normal survey operations.
This activity, sometimes referred to as ``flickering'', is a characteristic of LBVs \citep{Dorfi2003,doi:10.1080/21672857.2012.11519705}. 
The brightest and fastest of these outbursts occurred approximately 380 days after discovery on MJD 59845.48 (2022-09-23 11:31:12) with a peak absolute magnitude of $M_{\rm \wps}=-11.21\pm0.19$ and a decline rate of $dm_{\rm \wps}/dt=0.48\pm0.08$ measured over five days.
Pan-STARRS has since observed additional fainter and slower flickering episodes at 760$-$843\,days and 1,074$-$1,205\,days after discovery.
While not construed as a kilonova contaminant when considering the activity as a whole, the extreme short-scale flickering makes for a prominent contaminant if a single outburst is observed in isolation. 
The \wps\,filter light curve for the initial discovery outburst and first flickering episode of AT~2021yll is shown in Figure\,\ref{fig:2021ylllc} where the grey-shaded region highlights the kilonova-like fade of the brightest outburst. For completeness, we include the photometry measurements of the additional flickering episodes in the supplementary file (see section on Data Availability).

\subsubsection*{\textbf{AT~2021aanx}}
\begin{figure}
    \centering
    \includegraphics[width = 0.95\columnwidth]{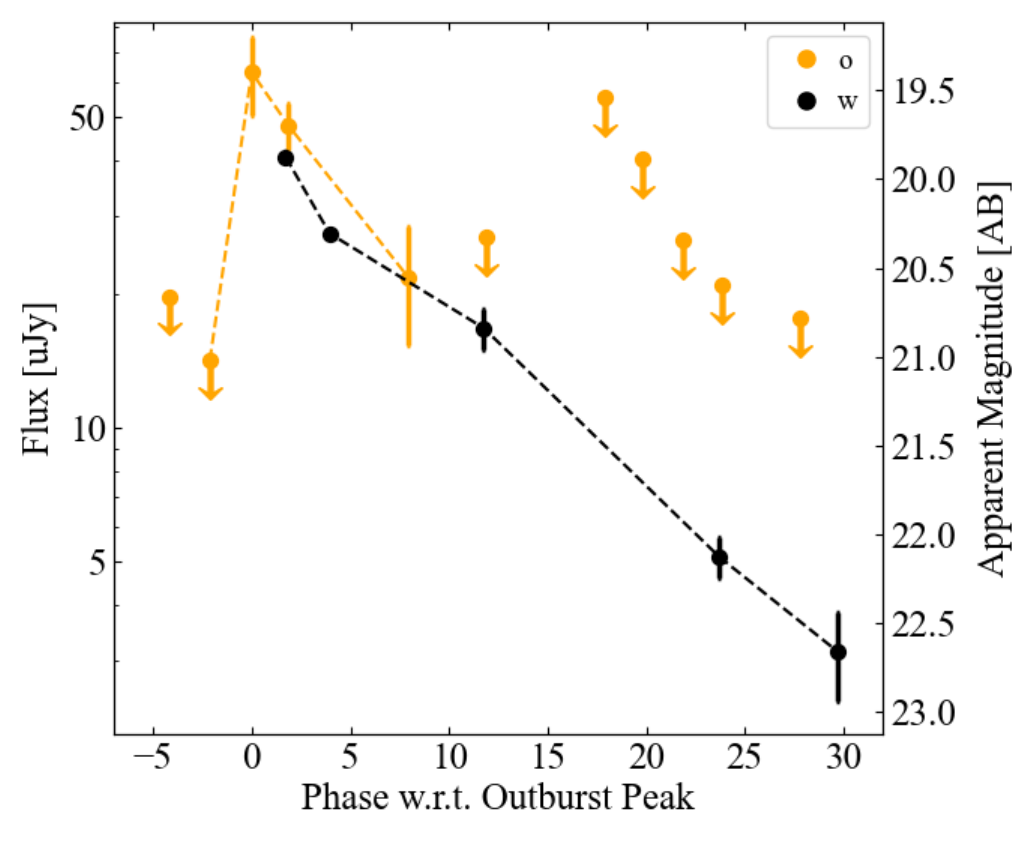}
    \includegraphics[width = 0.95\columnwidth]{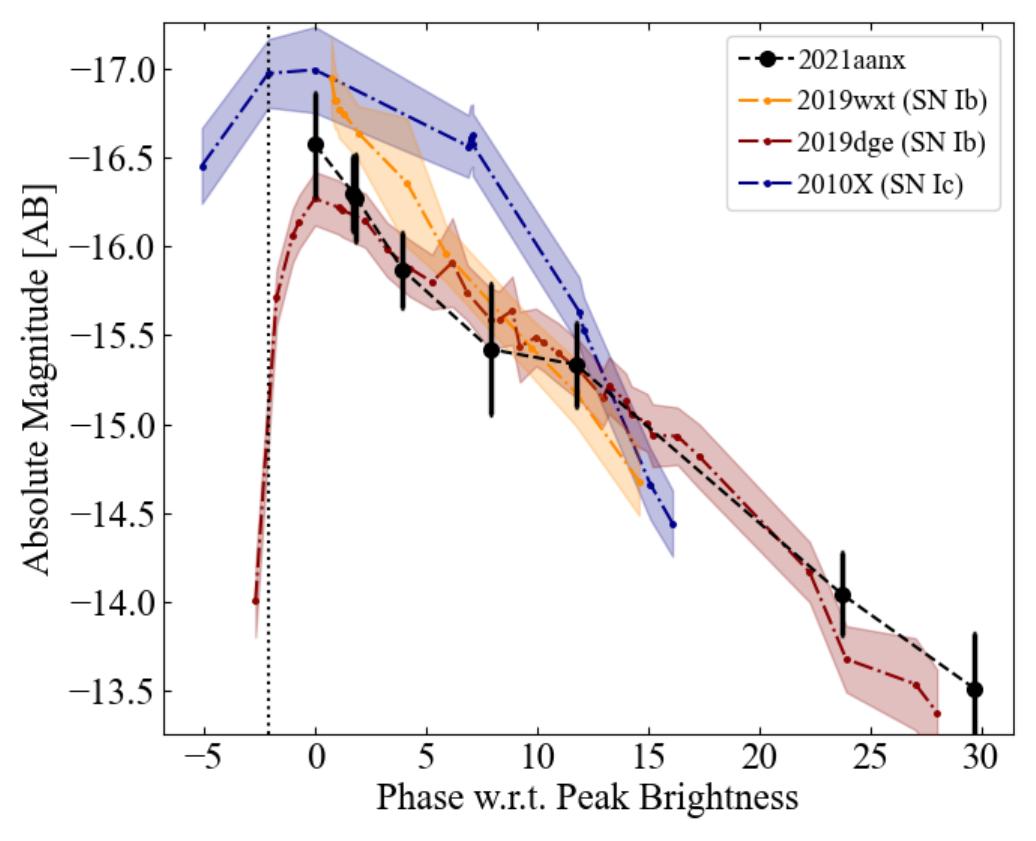}
    \caption{Top panel: The full ATLAS + Pan-STARRS light curve for AT~2021aanx. Data points with downward-pointing arrows represent limits at 3$\sigma$. The dashed lines highlight the fast-evolving trend. Bottom panel: The combined \wps + \oat\,filter light curve of AT~2021aanx plotted alongside the $r_{\rm SDSS}$\,filter light curves of ultra-stripped supernovae SN~2019wxt \citep{2023A&A...675A.201A}, SN~2019dge \citep{Yao_2020} and SN~2010X \citep{2010ApJ...723L..98K} in absolute magnitudes. The shaded colour regions denote the error space in the measurements of the compared supernovae. The vertical dotted black line represents the time of the latest ATLAS non-detection before the Pan-STARRS discovery of AT~2021aanx.}
    \label{fig:2021aanxlc}
\end{figure}

AT~2021aanx was discovered on MJD 59491.253 (2021-10-04 06:04:19 UTC) at a peak apparent magnitude of $m_{\rm \wps}=19.87\pm0.03$ in spiral host galaxy KUG 2316+183 which is at a luminosity distance of $D_{\rm d}=164\pm17$\,Mpc ($\mu=36.07\pm0.21$; from NED), implying a peak absolute magnitude of $M_{\rm \wps}=-16.30\pm0.21$. 
Follow-up on MJD 59493.495 (2021-10-06 11:52:48 UTC) observed a fade of 0.43\,magnitudes, hence a decline rate of $dm_{\rm \wps}/dt=0.19\pm0.02$. 
While the transient sits above the plausible kilonova contours in the $M-\dot{M}$ locus plot in Figure\,\ref{fig:fastfinder}, the ATLAS forced photometry suggests an unusually fast rise. 
It constrains the rise time within three days of the ATLAS peak measured in the \oat\,filter. 
Hence, we triggered additional Pan-STARRS follow-up to quantify the source. 
This revealed the decline rate had lessened, and the transient remained visible for 28 days, confirming it was not a kilonova candidate. 
The combined ATLAS and Pan-STARRS photometry is shown in the top panel of Figure\,\ref{fig:2021aanxlc}. 
While AT~2021aanx lies within the SN~IIb shock cooling regime of the $M-\dot{M}$ locus plot, the extended decline with no clear evidence of a rise into the main supernova event suggests that a USSN origin is more plausible. USSNe with rise times less than one week and absolute magnitudes ranging from $-12$ to $-17$ in the optical filters have been theorised \citep{10.1093/mnras/stv990} and previously documented in events such as the Helium-rich Type Ib events SN~2019wxt \citep{2023A&A...675A.201A} and SN~2019dge \citep{Yao_2020}, and the Helium-poor Type Ic SN~2010X \citep{2010ApJ...723L..98K}. 
Considering that the \wps\,filter follows the $r_{\rm SDSS}$\,filter and the Pan-STARRS \rps\,filter to a good approximation, we compare the  $r_{\rm SDSS}$\,filter light curves of SN~2019wxt, SN~2019dge and SN~2010X, to the combined \wps + \oat\,filter light curve of AT~2021aanx in the bottom panel of Figure\,\ref{fig:2021aanxlc}. 
Here, we applied an offset to match the \wps\,filter flux to the \oat\,filter flux, using the epochs measured $\approx1.8$\,days after the ATLAS peak as the reference point, so that a single trend line can represent the light curve of AT~2021aanx. 
Effectively, we are measuring the $\wps-o_{\rm A}$ colour at one epoch and applying that to the rest of the data. 
In reality, this will change as the spectral profile changes. 
Still, it is a reasonable approximation given the short timescales over which these offsets were applied and the close relationship between the \wps\,filter and \oat\,filter \citep{2012ApJ...750...99T,2018PASP..130f4505T}. 
We find that the light curve of AT~2021aanx is analogous to that of the fainter and faster Helium-rich SN~2019dge, and SN~2019wxt to a lesser extent.

\subsubsection*{\textbf{AT~2021aeba}}
Pan-STARRS discovered AT~2021aeba on MJD 59526.57 (2021-11-08 13:40:48 UTC) at a peak apparent magnitude of $m_{\rm \wps}=21.69\pm0.08$ in spiral host galaxy SDSS J080844.91+184204.7 which is at a luminosity distance of $D_{\rm L}=195\pm14$\,Mpc ($\mu=36.45\pm0.17$), implying a peak absolute magnitude of $M_{\wps}=-14.87\pm0.19$. 
Follow-up was conducted two days later, measuring a fade of $dm_{\rm \wps}/dt=0.31\pm0.10$ and we found no historic activity in Pan-STARRS or ATLAS.
However, \cite{2021TNSAN.293....1A} showed that variable flux behaviour before the Pan-STARRS discovery existed in forced photometry in the ZTF survey, which lasted 70+ days.
Additional Pan-STARRS observations four days after the Pan-STARRS discovery displayed a decrease in the decline rate and no significant colour evolution, all of which indicates an LBV origin.

\subsubsection*{\textbf{AT~2022kjm}}
\begin{figure}
    \centering
    \includegraphics[width = 0.95\columnwidth]{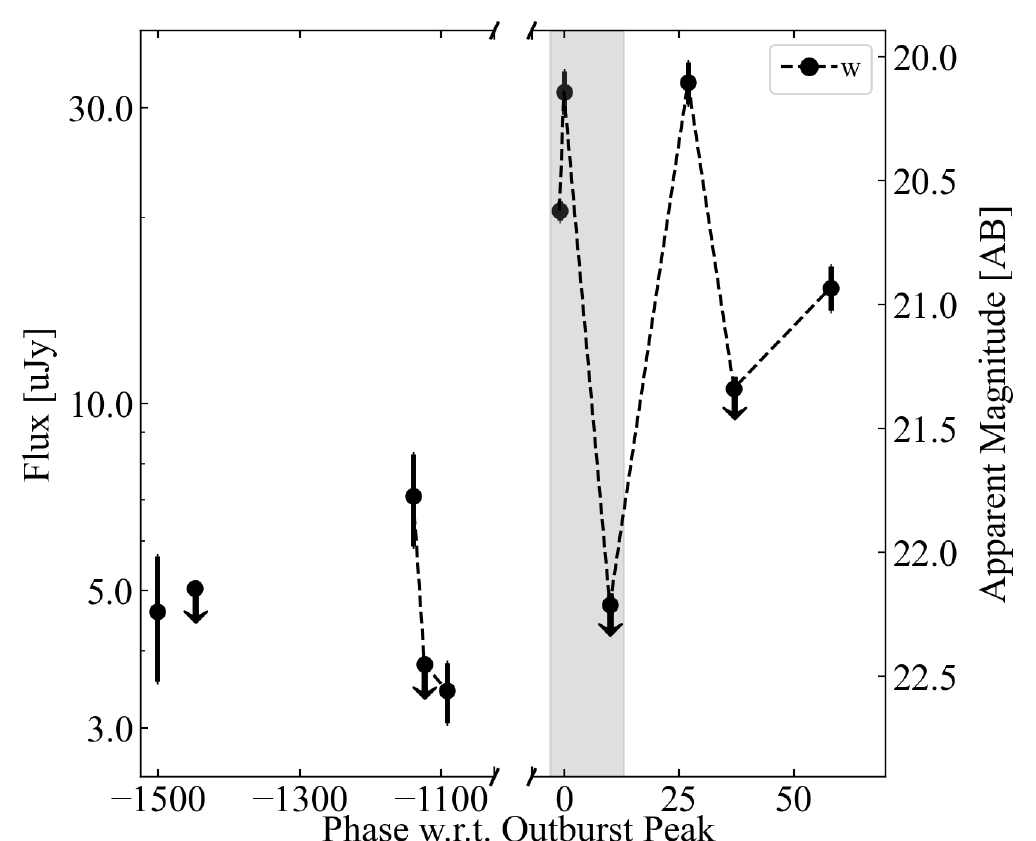}
    \caption{The Pan-STARRS \wps\,filter light curve for AT~2022kjm displaying the fast outburst measured at discovery (shaded in grey) and the low-level activity observed three years prior. Data points with downward-pointing arrows represent limits at 3$\sigma$. The fast outburst is a compelling kilonova contaminant when isolated.}
    \label{fig:2022kjmlc}
\end{figure}

Pan-STARRS discovered AT~2022kjm on MJD 59721.309 (2022-05-22 07:24:58 UTC) at a magnitude of $\wps=20.62\pm0.04$ in the spiral host galaxy IC 4177, which is at a luminosity distance of $D_{\rm L}=36\pm3$\,Mpc ($\mu=32.75\pm0.17$). 
One day later, the transient had brightened to an apparent magnitude of $\wps=20.14\pm0.08$, hence an absolute magnitude of $M_{\wps}=-12.72\pm0.19$.
Follow-up in the \grips\,filters ten days later had shown AT~2022kjm to have faded from view, implying a decline rate of $dm_{\wps}/dt=0.21\pm0.06$. 
However, Pan-STARRS forced photometry revealed multiple minor oscillations $\geq$3 years prior, and another outburst approximately one month later was observed through normal survey operations. 
A further outburst at $\wps\simeq21.4$ was observed with Pan-STARRS starting nearly one year later (MJD 60024), strongly suggesting an LBV origin. 
The Pan-STARRS light curve of AT~2022kjm (during the first 2022 outburst) is shown in Figure\,\ref{fig:2022kjmlc}. 

\subsubsection*{\textbf{SN~2022abom}}
Pan-STARRS discovered SN~2022abom on MJD 59907.434 (2022-11-24 10:24:58 UTC) at an apparent magnitude of $m_{\rm \wps}=20.51\pm0.04$ in the elliptical host galaxy WISEA J034109.82-025031.7, which is at a luminosity distance of $D_{\rm L}=158\pm11$\,Mpc ($\mu=35.99\pm0.17$), implying an absolute magnitude of $M_{\rm \wps}=-15.69\pm0.18$.
Follow-up two days later showed the transient had faded by 0.8 magnitudes, signifying a decline rate of $dm_{\rm \wps}/dt=0.39\pm0.05$. 
As the fast-fading or flickering LBV candidates are exclusively in blue star-forming hosts, this fast and faint source in a passive galaxy caught our attention. 
However, additional observations the following night showed the transient had rebrightened to a luminosity similar to that at the time of discovery, and a spectrum taken on MJD 59913.204 (2022-11-30 04:53:35 UTC), along with further follow-up, revealed it was a sub-luminous 02es-like SN Ia \citep{2022TNSCR3508....1G}. 
The initial Pan-STARRS photometry thus indicates a prominent, short-lived and rapidly  evolving early excess feature that has been detected in a handful of other 02es-like SNe Ia \citep[e.g.][]{cao2015strong,2021ApJ...919..142B,Srivastav_2023,2024MNRAS.527.9957X}. These peculiar 02es-like SNe Ia thus represent another type of contaminant for kilonovae identified for future versions of our kilonova prediction algorithm. A detailed analysis of SN~2022abom involving Pan-STARRS data is in preparation (Srivastav et al. 2025).

\section{Discussion of LBV contaminants in future kilonova searches}
\label{sec:disc}
Through combining photometric observations across multiple surveys, we conclude that none of our most promising transients were viable kilonova candidates and instead were either LBV outbursts, Galactic CVs or had supernova-like origins.

Within the sample of 11 fast-evolving transients, we had two spectroscopically confirmed LBVs (AT~2017dau and SN~2021qvw) and another four plausible LBV candidates as deduced from their photometric behaviour (AT~2017des, AT~2021yll, AT~2021aeba and AT~2022kjm). 
This makes LBVs the most prominent contaminant identified in our Pan-STARRS search for kilonovae.
The occurrence rate of LBV outbursts as a whole is not well understood. 
Therefore, it's crucial to discern the intrinsic rate of LBV outbursts to assist in contamination rates in future kilonova searches. 

\subsection{The rate of LBV outbursts from the ATLAS survey}
\label{sec:LBVrates}

The sporadic nature of the Pan-STARRS cadence, its adjustment to follow NEO discoveries and the fill factor of the camera (see Section\,\ref{subsec:observe-strat}) all conspire to mean that the survey is not ideally suited to determine the rates of fast transients and LBVs. 
However, in parallel, we have been running a more complete and volume-limited survey for transients with ATLAS \citep{2020PASP..132h5002S}.
Due to their low luminosity, LBV outbursts are found only within a few tens of Mpc with reasonable completeness in ATLAS, reducing the numbers discovered. 
The ATLAS survey typically has a consistent 1-4 day cadence (with the variation typically due to weather). 
While it is shallower, it provides a much higher degree of sampling to allow such local rates to be calculated with better precision. 

We have constructed a $3.5$\,year-long sample (21st September 2017 - 21st March 2021) of transients observed by the ATLAS survey within a distance of $100$\,Mpc. 
We will be presenting a series of papers describing the methodology and results of this ``ATLAS Local Volume Survey'' in full, which has already been discussed to some extent in \cite{2020PASP..132h5002S} and \cite{2022MNRAS.511.2708S}. 
Therefore, this analysis only presents the details necessary for determining LBV rates. In the ATLAS sample of 902 transient events, 11 were spectroscopically classified as LBV outbursts. Of these, nine had outbursts whose peak absolute magnitudes reached $M_{\rm peak}< -11$, i.e. the range at which kilonovae are expected to be found.
All nine of these  LBVs in our ATLAS sample were found within $\sim50$\,Mpc, with our brightest LBV outburst also being the most distant event with a spectroscopic redshift of $z=0.0125$ \citep{2020TNSCR2126....1D} or a luminosity distance of $\approx54$\,Mpc. 
However, there is a very large difference in the recovery efficiency $\eta$ between the faint ($M_{\rm peak}\simeq-11.8$) and bright ($M_{\rm peak}\simeq-15$) LBV events due to their diverse light curves.

It is clear from our Pan-STARRS search that any wide-field optical survey aiming to detect kilonovae will detect the contaminants described in Section\,\ref{sec:results} and will have to filter them out efficiently. 
One additional question is how such faint and fast-evolving contaminants may affect searches for kilonova when there is also a gravitational wave trigger, and both the 2-dimensional sky map and 3-dimensional volume are large.
For this illustrative calculation, we focus on the maximum rate of LBV outbursts to determine if they could be a contaminant in future searches for kilonovae from gravitational wave triggers. 
Correcting for the volume of sky observed by the two ATLAS units (88\% of the full celestial sky), we calculate ``maximum'' and ``minimum'' limits on the LBV volumetric rates using the faint LBV and bright LBV recovery efficiencies on the full sample. 
See Appendix~\ref{sec:Appendix ATLAS} for further details on the ATLAS survey, our methodology for calculating LBV volumetric rates, and determining the recovery efficiencies.
These LBV rates are reported in Table\,\ref{tab:rates} along with the rates of other transients for comparison.
The LBV rates are dominated by the recovery efficiency, which is dependent on the luminosity function of LBVs.
With only 11 sources, the luminosity function is uncertain. 
Still, the numbers are sufficient to highlight that the rate of LBV outbursts that are brighter than $M_{\rm peak} < -11$ is a few times the core-collapse supernova rate. 
This is corroborated just by counting the number of eruptive or explosive transients discovered within $D\lesssim10-20$\,Mpc in any survey, indicating the numbers are dominated by transients that are fainter than supernovae. 
A combination of LRNe, LBV outbursts and ILRTs are the most numerous, as illustrated in Table\,\ref{tab:rates} and it can be difficult to classify them as independent classes of objects securely. 

\begin{table}
\centering
\begin{tabular}{lll}
\hline
Object        & Rate and error               & Source     \\
              & $\mathrm{Gpc^{-3}\,yr^{-1}}$  &     \\\hline
LBVs (faint)   &    $60~ \pm 30~ \times 10^{4}$    &   ATLAS survey - this paper  \\
LBVs (bright)  &    $1~~~\pm 0.4 \times 10^{4}$  &   ATLAS survey - this paper   \\
CCSNe          &    $10~\pm 5~~~ \times 10^{4}$    & ZTF: \cite{2020ApJ...904...35P}    \\
Ia SNe    &    $2.4 \pm 0.2\times 10^{4}$    &  ZTF: \cite{2020ApJ...904...35P}     \\
LRNe          &   $8~~~ \pm 6~~~  \times10^{4}$ & ZTF:  \cite{2023ApJ...948..137K} \\
IRLTs          & $0.3 \pm 0.2~~ \times10^{4}$ & ZTF: \cite{2023ApJ...948..137K}\\
Optically Bright  KNe   &     $<0.09      \times 10^{4}$ & ZTF: \cite{2021ApJ...918...63A}  \\
BNS mergers   &    $0.01 - 0.17\times 10^{4}$      & LVC:  \cite{2023PhRvX..13a1048A}  \\
\hline
\end{tabular}
\caption{Our estimated rate of LBV outbursts in the Local Universe is listed alongside the rates for other well-known classes of optical transients.
The binary neutron star merger rate from the LIGO-Virgo collaboration is listed for comparison.}
\label{tab:rates}
\end{table}

\subsection{The LBV contamination rate in future joint gravitational wave and electromagnetic searches}

The results of our observational survey with Pan-STARRS indicate that we can find fast-evolving and faint transients that occupy the area of the $M-\dot{M}$ locus plot in which we expect to find kilonovae. 
The lack of discovery will, in principle, offer an estimate of the upper limit on the kilonova rate. 
This is challenging with both Pan-STARRS telescopes due to the non-uniformity of the observing cadence and the fill factor of the giga-pixel cameras and is beyond the scope of this initial paper. 
The main result of this search highlights that the true rate of contaminants far outweighs kilonovae in such searches with optical data only (by one to two orders of magnitude). 
In this section, we comment on the implications of these contaminants when searching the gravitational wave skymaps produced by the LVK detectors for optical counterparts to BNS mergers. 

Over the entirety of the third LIGO-Virgo science run (O3), 14 GW triggers were issued in real time for BNS and neutron star-black hole (NSBH) mergers. 
These have not been formally retracted, but in reality, only one BNS merger event \citep[GW190425;][]{190425} survived offline analysis. 
Three NSBH events were likely true signals, with two having fairly confident NS masses for the secondary \citep{NSBH} and the third having a secondary mass of ambiguous nature \citep[GW190814;][]{190814}. 
The median size of the skymaps for these GW triggers (4 real and 10 likely bogus) were $633$\,deg$^{2}$ and $3646$\,deg$^{2}$ for the 50\% and 90\% confidence areas, respectively. 
The crucial component of searching these skymaps was removing the contaminating supernova and transient population. 
The initial step is to use the GW trigger's date and time to reject transients that are spatially coincident but had some flux in pre-existing survey data before the GW detection time.
Searching GW skymaps of several hundred square degrees will find mostly old supernovae that have exploded weeks or months prior. 
\cite{2016MNRAS.462.4094S} predict that surveys with limiting magnitudes $m=21$ will typically find five new supernovae events per $100$\,deg$^{2}$ within a 10-day period.

A more relevant question is how many kilonova impostors are expected to be detected in a typical GW skymap? 
Even now, nearly eight years after the discovery of AT~2017gfo as the kilonova associated with the GW170817, we have had minimal experience in real-time searching of an error map of a few hundred square degrees of a real BNS merger. 
The skymap of GW190425 was 8284 deg$^{2}$ \citep{190425} due to having signal in only one detector, which made optical searches challenging \citep{2019ApJ...885L..19C,190425_panstarrs} and approximately half the map was in the daytime sky.

Photometric follow-up can be a quick and effective method for ruling out any new supernova event through the lack of rapid light curve evolution.
However, identifying kilonova impostors, such as rapidly changing LBV outbursts or shock cooling (as shown in Section\,\ref{sec:results}) will require follow-up observations to track light curve evolution accurately. 
During the fifth observing run of LVK (O5), we may expect to typically search sky areas of 100-500 square degrees for GW detections at luminosity distances of 200-400\,Mpc \citep{2020LRR....23....3A}. 
We also may expect to use the Rubin Observatory for deep and sensitive searches within its Legacy Survey of Space and Time \citep{2019ApJ...873..111I}, as described in \cite{2022ApJS..260...18A} and more recently in \cite{2024arXiv241104793A}. 

\begin{figure}
    \centering
    \includegraphics[width = 0.95\columnwidth]{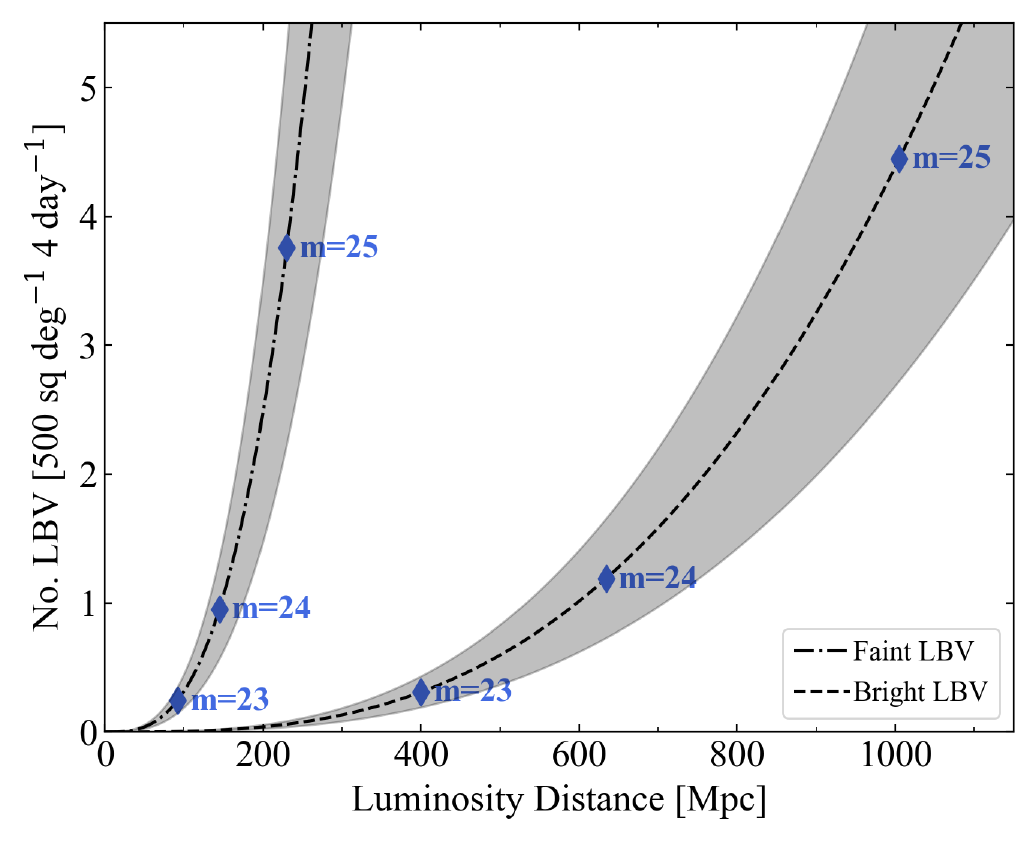}
    \caption{The predicted number of LBV events per $500$\,deg$^{2}$ per 4-day period, determined from the maximum and minimum volumetric rates depicted in Table\,\ref{tab:rates}. The dotted-dashed line represents the higher rate of faint luminosity outbursts ($M_{\rm peak}=-11.8$), while the dashed-dashed line represents the lower rate of bright outbursts ($M_{\rm peak}=-15$). The shaded regions represent the errors on each rate. The blue labels depict the apparent magnitude limits for faint and bright LBV events along each trend, indicating how many events are expected at their maximum observable luminosity distances.}
    \label{fig:skymap_contam}
\end{figure}

Using our volumetric rates listed in Table\,\ref{tab:rates}, we plot the number of LBV events we expect to find within an arbitrary 500 square degree tile of night sky over a 4-day period out to a distance of 1 Gpc in Figure\,\ref{fig:skymap_contam}. 
The blue labels indicate the apparent magnitude limit for faint ($M_{\rm peak}=-11.8$) and bright ($M_{\rm peak}=-15$) LBV outbursts at the corresponding distances of their respective trends. 
If we can date the explosion epoch of the outbursts with an uncertainty of four days, then the numbers in this figure are a reasonable estimate of the rate of contaminating sources. 

As described in \cite{2020LRR....23....3A}, we may expect a large number of 100-500 square degree skymaps, and  \cite{2024arXiv241104793A} proposes trigger strategies based on such events.
Therefore, we estimate that a search with LSST covering a sky map size of 500 square degrees over four days, which reaches limiting magnitudes of $m\approx25$ \citep[the Silver event observing strategy described in][]{2024arXiv241104793A}, will likely pick up around $4\pm2$ faint LBV-type outbursts within 200-400 Mpc, and a slightly larger number of $5\pm1$ for the brighter (but rarer) LBV events out to 1 Gpc. 
These numbers imply that we will not be dominated by such LBV contaminants in searches for kilonovae in O5. 
However, \cite{2024arXiv241104793A} also notes that due to the detector duty cycles, O5 may discover a substantial number of mergers which are one or two detector events only, whose skymaps are of order 1000 square degrees or more. 
In that case, our rate estimates indicate that LBV-type outbursts may be significant contaminants. 
Such contaminants that do appear must be identified and ruled out quickly so that vital follow-up time on $8-10$\,m telescopes (for reliable spectroscopy at magnitudes $m\leq23$) is not wasted by observing them. 
The LSST Wide-Fast-Deep survey history will likely play an important role in finding activity before the GW event trigger, particularly through forced photometry at the candidates' positions.

\begin{figure}
    \centering
    \includegraphics[width = 0.95\columnwidth]{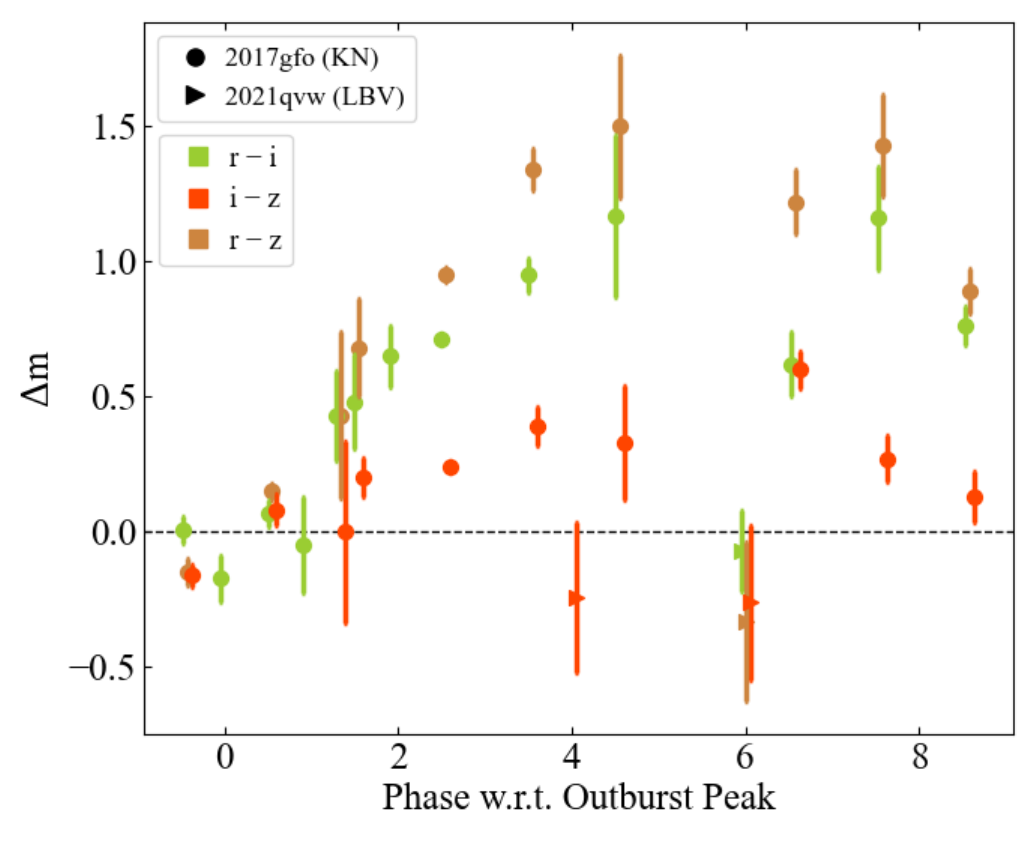}
    \caption{Comparison of the $r-i$,\,$i-z$\,and $r-z$ colour evolution between known kilonova AT~2017gfo and the LBV precursor to SN~2021qvw identified in the Pan-STARRS sample of fast-evolving transients. The phase is with respect to the outburst peak brightness for the LBV precursor. There is a significant difference in the $r-i$\,and $r-z$ colours between the two objects at times $\geq6$\,days after peak, where LBVs appear to remain bluer in colour.}
    \label{fig:colour_comp}
\end{figure}

We demonstrated in Section \ref{sec:results} that continued imaging can help distinguish LBVs, along with the fact that they occur almost exclusively in star-forming regions of late-type galaxies. 
Photometric colour evolution in the optical, near-infrared, and now mid-infrared with the James Webb Space Telescope \citep[see the recent detections of kilonova in][]{2024Natur.626..737L} is a further way to distinguish kilonovae from LBVs. 
In Figure\,\ref{fig:colour_comp} we present the $r-i$,\,$i-z$\,and $r-z$ colour evolution for kilonova AT~2017gfo and the LBV precursor to SN~2021qvw confidently identified in the Pan-STARRS sample of fast-evolving transients. 
We find that at $\geq 6$ days after peak brightness, there is a notable difference in the $r-i$\,and $r-z$ colours between AT~2017gfo and the LBV precursor. 
Unlike what is expected for a kilonova, the LBV precursor shows no significant colour change to a red excess at late times.
Therefore, eliminating such contaminants is feasible through photometric observations across multiple filters, particularly \ensuremath{riz} filters, which are more easily obtained than spectra at faint magnitudes.

\section{Conclusions}
\label{sec:conclusions}

We explored 3.14 years of Pan-STARRS survey data searching for rapidly evolving transients within a luminosity distance $D_{\rm L}\leq\dlim$\,Mpc that could plausibly be considered kilonova candidates based on their photometric evolution through our kilonova prediction algorithm. 
We employed forced PSF photometry and nightly flux stacking and combined them with data from the ATLAS and ZTF surveys to improve our ability to recognise kilonova candidates. 
Out of the 1,852 transient sources identified and associated with a host galaxy within $\leq\dlim$\,Mpc,\,\,577 of these were Pan-STARRS discoveries. 
A subset of 175 events met the luminosity criterion for kilonovae with discovery absolute magnitude $M>-16.5$, of which 11 were highlighted as fast-fading.

None of these most promising transients turned out to be viable kilonova candidates, instead turning out to be either foreground Galactic stellar outbursts, LBV outbursts, shock cooling signatures or low luminosity SN-like events. 
Our fast-evolvers sample was primarily contaminated by outbursts from LBVs, accounting for 55\% of the sample. 
Using data from the ATLAS survey, we calculate maximum and minimum volumetric rates for faint and bright LBV outbursts as $R_{\rm Max} = 6 \pm 3 \times 10^{5}\,\mathrm{Gpc^{-3}\,yr^{-1}}$ for the faint $M_{\rm peak}\simeq-11.8$ sample and $R_{\rm Min} = 1 \pm 0.4 \times 10^{4}\,\mathrm{Gpc^{-3}\,yr^{-1}}$, for the brighter $M_{\rm peak}\simeq-15$ sample. 
We converted these volumetric rates into an expected contamination rate for LSST-like surveys searching EM counterparts in LVK GW skymaps. 
Surveys reaching limiting magnitudes of $m\approx25$ would expect to find 4$\pm2$ faint LBV outbursts per $500$\,deg$^{2}$ within a 4-day window within the typical distance that BNS mergers would be found in O5 (200-400 Mpc). 
Such a sky map may be typical in O5, depending on the sensitivity and duty cycle of a three- or four-detector network.  
While they are not an overwhelming contaminant source, they must be considered as possible interlopers when only photometric information is available in the early part of kilonova searches. We also conclude that with both pre-discovery forced photometry and follow-up photometry, they can be distinguished from kilonovae on rapid timescales.  

\section*{Acknowledgements}
Pan-STARRS is a project of the Institute for Astronomy of the University of Hawaii, and is supported by the NASA SSO Near Earth Observation Program under grants 80NSSC18K0971, NNX14AM74G, NNX12AR65G, NNX13AQ47G, NNX08AR22G, 80NSSC21K1572 and by the State of Hawaii.
The Pan-STARRS1 Sky Survey data were facilitated by  the University of Hawaii, the Pan-STARRS Project Office, the Max Planck Society 
(MPIA, MPE), Johns Hopkins University, Durham University,  University of Edinburgh,  Queen's University Belfast, the Harvard-Smithsonian CfA the Las Cumbres Observatory Global Telescope Network Incorporated, the National Central University of Taiwan, the Space Telescope Science Institute, NASA Grant No. NNX08AR22G, 
NSF Grant No. AST-1238877, the University of Maryland, Eotvos Lorand University, and the Los Alamos National Laboratory. 
The Asteroid Terrestrial-impact Last Alert System (ATLAS) project is primarily funded to search for Near-Earth asteroids through NASA grants NN12AR55G, 80NSSC18K0284, and 80NSSC18K1575; byproducts of the NEO search include images and catalogs from the survey area. This work was partially funded by Kepler/K2 grant J1944/80NSSC19K0112 and HST GO-15889, and STFC grants ST/T000198/1 and ST/S006109/1. The ATLAS science products have been made possible through the contributions of the University of Hawaii Institute for Astronomy, the Queen's University Belfast, the Space Telescope Science Institute, the South African Astronomical Observatory, and The Millennium Institute of Astrophysics (MAS), Chile. 
The ZTF forced-photometry service was funded under the Heising-Simons Foundation grant \#12540303 (PI: Graham).
S. J. Smartt and K. W. Smith acknowledges funding from STFC Grants ST/Y001605/1, a Royal Society Research Professorship and the Hintze Charitable Foundation.  
S. Srivastav and  D. R. Young acknowledge funding from STFC Grants ST/Y001605/1, ST/X001253/1, ST/X006506/1, ST/T000198/1. 
M. Nicholl is supported by the European Research Council (ERC) under the European Union’s Horizon 2020 research and innovation programme (grant agreement No.~948381) and by a Fellowship from the Alan Turing Institute.
We acknowledge use of NASA/IPAC Extragalactic Database (NED),  operated by JPL, Caltech under contract with NASA. 
%

\section*{Data Availability}
\label{sec:dataAvailability}
A CSV file has been provided, containing all the Pan-STARRS and ATLAS optical forced photometry of the 11 fast-evolving transients behind Figures\,\ref{fig:outburstlcs}\,through to\,\ref{fig:2022kjmlc} and Table\,\ref{tab:fastobjs} in a machine-readable format. 
Measurements are in AB magnitudes and microJansky flux, representing nightly stacked fluxes from individual difference images.  
The total exposure time for each night is provided, which is typically the sum of sub-exposures combined. 
All photometry are uncorrected for galactic and host galaxy dust extinction. 
Magnitude limits are quoted to 3$\sigma$.
Where a \wps filter exposure did not exist, a modelled \wps filter ``mod-w'' was constructed, if available, from \grips filter exposures taken on the same night. 
The phase is with respect to the peak brightness of the associated outbursts outlined in Table\,\ref{tab:fastobjs} and Section\ref{sec:results}.


\bibliographystyle{mnras}
\bibliography{main} 




\appendix
\section{The rate of LBV outbursts from the ATLAS survey}
\label{sec:Appendix ATLAS}
On a typical night, two ATLAS units operating in tandem, taking $4\times30$\,second dithered exposures in coordinated quadrants, will detect around 10-15 extragalactic transient candidate sources after processing, filtering and the removal of known sources (stars, AGN, movers, etc.,). 
A human scanner later identifies these good candidates in the nightly data and registers them on the TNS. 
Similar to our Pan-STARRS search, a selection of these good candidates was linked to galaxies with a known redshift of $z\lesssim0.024$ using the Sherlock algorithm, corresponding to a luminosity distance of $D_{\rm L}\lesssim100$\,Mpc. 
The radius of association with galaxies of known redshift was set at 50\,kpc, and all transients detected at such extreme radial distances underwent a thorough vetting process to eliminate any background supernovae. 
This is achieved by carefully inspecting the Pan-STARRS and SDSS images for more credible and distant hosts coupled with their spectroscopic classifications, if available. 
The sample also includes a subset of confirmed supernovae whose host galaxy redshift was unknown, but the classification redshift reported on the TNS was within our distance cutoff. 
The sample consists of 902 transients in total, of which 134 were without a spectroscopic classification. 

We find 11 transients spectroscopically classified as being LBVs within the sample. 
Of these, nine had outbursts whose peak absolute magnitudes reached that expected for kilonovae (i.e., $M_{\rm peak}<-11$). 
We did not search for any LBV candidates in the unclassified subset, as the wide variety in the types of ``Gap transients'' (those brighter than classical novae but fainter than normal supernovae; \citealt{2019NatAs...3..676P}) and lack of colour information in ATLAS observations make it difficult to distinguish an LBV from other Gap transients without the aid of spectral clues. 
The best $5\sigma$ limiting magnitude ATLAS can achieve in a 30-second exposure during dark time is $m_{\rm 5\sigma}=19.8$. 
For more typical observations, this limit is shallower at $19.2\leq m_{\rm 5\sigma}\leq19.5$ \citep{2018PASP..130f4505T}. 
This constrains the distance ATLAS can detect the brightest of LBV outbursts to $<70$\,Mpc ($z<0.016$), a distance range favoured by classification surveys, so it is doubtful the unclassified subset will contain LBVs.

From the ATLAS sample, we can define the volumetric occurrence rate of any given transient type as
\begin{equation}
R = \frac{N}{\epsilon VT}
\end{equation}
Here, $N$ is the number of transients observed by ATLAS over a time $T$ within a volume $V$. The efficiency factor $\epsilon$ represents the fraction of transients that are detected and eventually registered by human scanners. 
$\epsilon$ accounts for inefficiencies and systematic limitations, such as the intrinsic luminosity of the transient type, the lifetime (light curve length) of the transient type, sky coverage, breaks in observation due to weather patterns and solar conjunction, obscuring by the galactic plane and foreground Milky Way extinction outlined in the \citet{2011ApJ...737..103S} dust maps, variations in sensitivity and background noise, image subtraction artefacts, the competence of the machine learning algorithm employed by the ATLAS transient server to identify real transients during the filtering process, and the element of human error during the scanning process. 
Consequently, the efficiency factor is not a constant value. 
It will vary as a function of $V$ (and to some extent $T$), where brighter and longer-lived transient events will be more readily detected.

Therefore, to estimate the ATLAS efficiency factor for LBV events (at the faint and bright ends of the distribution), we used the ATLAS efficiency simulation tool \citep{2021PhDOwen}. 
The tool generates a population of 10,000 simulated light curves, randomly distributed across a defined time frame, redshift, and sky coordinate range for a given input light curve. 
It then assesses the recovery rate based on the history of ATLAS observations. 
We set the ATLAS survey declination limits to $-50^{\circ} \leq \delta \leq 90^{\circ}$, i.e. the sky visible from Mauna Loa and Haleakala on which the ATLAS units, used in the ATLAS Local Volume Survey, are based. 
No right ascension limits are set since an event in solar conjunction may still be detectable when it emerges. 

The ATLAS transient server, as described in the paper by \cite{2020PASP..132h5002S}, has specific criteria to identify real astrophysical transients. 
To be considered a real transient, a minimum of three co-spatial detections with a significance of $5\sigma$ are necessary from the four exposures taken on any given night. 
However, if a candidate has only three detections at the $5\sigma$ limit, it will likely be put on hold and not immediately promoted to the TNS by the human scanner. 
Subsequent detections on the following nights would increase the likelihood of promotion and registration on the TNS. To replicate this selection process and ensure that the simulated transient is detected on at least two nights, we require a minimum of seven detections for a simulated transient to be successfully recovered.

\begin{figure}
    \centering
    \includegraphics[width = 0.95\columnwidth]{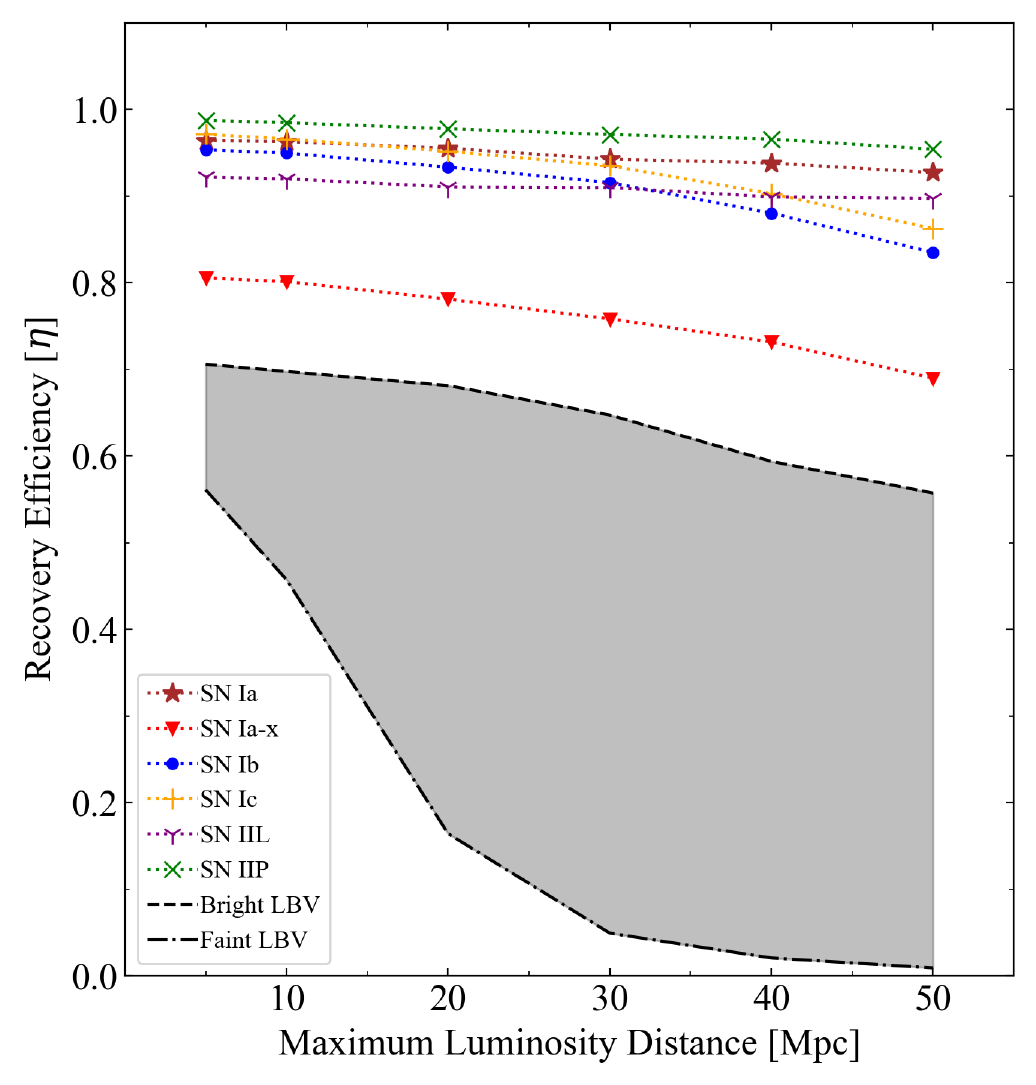}
    \caption{Simulated ATLAS recovery efficiency ($\eta$) as a function of luminosity distance for input light curves of SN~2020acma (SN~Ia), SN~2020hvp (SN~Ib), SN~2019yz (SN~Ic), SN~2021yja (SN~IIP), SN~2020adfj (SN IIL), SN~2019muj (intermediate SN~Ia-x), and AT~2018kle and AT~2020agp for the bright and faint LBVs, respectively. The grey-shaded region enclosed by the LBV trends represents the range in recovery efficiency across the various amplitudes of LBV outbursts.}
    \label{fig:eff_recov}
\end{figure}

In Figure\,\ref{fig:eff_recov}, we present the ATLAS recovery efficiency $\eta$ for different supernova types and LBVs from our ATLAS sample. In this context, recovery efficiency refers to the fraction of recovered events to the total number of simulated events within the volume enclosed by various luminosity distances rather than the efficiency at those specific distances. Hence, it can be said that the efficiency factor and recovery efficiency ($\epsilon$ and $\eta$, respectively) are equivalent for a fixed volume of space enclosed by the maximum luminosity distance at that value of $\eta$.

As expected, the simulations suggest consistently high recovery efficiencies of $\eta\geq0.8$ for the more commonly occurring supernova types (SN~Ia, SN~Ibc, and SN~II), exploding at any time up to a maximum distance of 50 Mpc. 
The intermediate luminosity SN~Iax ($M_{\rm peak}\approx-16.2$) used in this simulation, although starts with a lower recovery efficiency than the normal supernovae, remains mostly constant at $\eta\geq0.7$. 
The brightest LBV from our ATLAS sample, which has a peak absolute magnitude of $M_{\rm peak}=-15.0\pm0.1$, starts with an even lower recovery efficiency but behaves similarly to the SN~Iax by remaining roughly constant as the maximum distance increases. 
For a volume corresponding to a distance of 50\,Mpc, we obtain a recovery efficiency of $\eta=0.557$ for the bright LBV. 
However, the recovery efficiency for our faintest LBV, which has a peak absolute magnitude of $M_{\rm peak}=-11.8\pm0.1$, was considerably lower and fell off more steeply with increasing maximum distance. 
We measure a recovery efficiency of only $\eta=0.009$ for a volume corresponding to a distance of $50$\,Mpc. 
At larger distances, the recovery efficiency for the faint LBV effectively falls to zero, hence our cap of $50$\,Mpc for the maximum luminosity distance in our efficiency simulation, and, subsequently, the volume with which we calculate all volumetric rates. 


\bsp	
\label{lastpage}
\end{document}